\documentclass[prb,twocolumn,showpacs,superscriptaddress,floatfix]{revtex4}
\usepackage{graphicx,amsfonts,amssymb,amsmath}
\usepackage[pagebackref=false,colorlinks,linkcolor=blue,citecolor=red]{hyperref}

\def\be{\begin{equation}}
\def\ee{\end{equation}}
\def\bea{\begin{eqnarray}}
\def\eea{\end{eqnarray}}

 \newcommand{\ket}[1]{|#1\rangle}

 \newcommand{\comutator}[2]{[ #1,#2 ]}

\begin{document}

\title{Baxter-Wu model in a transverse magnetic field}

\author{Sylvain Capponi}
\email{capponi@irsamc.ups-tlse.fr}
\affiliation{Laboratoire de Physique Th\'eorique, Universit\'e de Toulouse and CNRS, UPS (IRSAMC), F-31062, Toulouse, France}

\author{Saeed S. Jahromi}
\email{s.jahromi@dena.kntu.ac.ir}
\affiliation{Department of Physics, K.N. Toosi University of Technology, P.O. Box 15875-4416, Tehran, Iran}
\affiliation{Lehrstuhl f\"{u}r Theoretische Physik I, Otto-Hahn-Stra{\ss}e 4, D-44221 Dortmund, Germany}

\author{Fabien Alet}
\email{alet@irsamc.ups-tlse.fr}
\affiliation{Laboratoire de Physique Th\'eorique, Universit\'e de Toulouse and CNRS, UPS (IRSAMC), F-31062, Toulouse, France}

\author{Kai Phillip Schmidt}
\email{kai.schmidt@tu-dortmund.de}
\affiliation{Lehrstuhl f\"{u}r Theoretische Physik I, Otto-Hahn-Stra{\ss}e 4, D-44221 Dortmund, Germany}

\begin{abstract}
We investigate the low-energy properties as well as quantum and thermal phase transitions of the Baxter-Wu model in a transverse magnetic field. Our study relies on stochastic series expansion quantum Monte Carlo and on series expansions about the low- and high-field limits at zero temperature using the quantum finite-lattice
method on the triangular lattice. The phase boundary consists of a second-order critical line in the 4-state Potts model universality class starting from the pure Baxter-Wu limit meeting a first-order line connected to the zero-temperature transition
point \mbox{($h\approx2.4$, $T=0$)}. Both lines merge at a tricritical
point approximatevely located at \mbox{($h\approx 2.3J$, $T\approx J$)}.
\end{abstract}
\maketitle

\section{Introduction}
\label{intro}

Spin models have played a major role since the early years of statistical mechanics, since they represent ideal systems to study the physics of classical and quantum phase 
transitions as well as associated critical behavior \cite{Stanley,sachdev}. The prototypical quantum spin model
with two-spin interactions is the Ising model in a transverse field (TFIM). It is (only) exactly solvable in one dimension at zero temperature or,
 equivalently, on the two-dimensional square lattice for a vanishing field. More generally, the $d$-dimensional TFIM at zero temperature can be mapped 
to the $(d+1)$-dimensional Ising model at zero field, both displaying a second-order phase transition in the $(d+1)$ Ising universality class which plays 
an important role for many other classical and quantum many-body systems \cite{Stanley}, e.~g.~in the duality to the $\mathcal{Z}_2$ gauge theory \cite{Fradkin79,Kogut79}
which has been discussed recently in the context of the topologically ordered toric code \cite{Kitaev03} in the presence of a magnetic field \cite{Trebst07,Hamma08,Vidal09a,Tupitsyn10,Dusuel11,Wu12}. 

The Ising model with three-spin interactions is an interesting system
with multiple spin interactions which was introduced by Baxter and Wu
on the triangular lattice in 1973 to describe ferrimagnetism and
critical phase transitions
\cite{baxter1,baxter2,baxter3}. Interestingly, this Baxter-Wu (BW)
model is exactly solvable. It displays a second-order thermal phase
transition in the 4-state Potts model universality class (without
logarithmic corrections). The latter has been confirmed by series
expansions \cite{BW_SE} and quantum Monte Carlo simulations
\cite{Penson,Santos,Novotny}. Recently, also the Baxter-Wu model in a
transverse magnetic field (BWTF) has been studied by series expansions
at zero temperature \cite{ssj}. Here clear signatures are found for a
first-order quantum phase transition separating the ordered phase from
the polarized phase present at large magnetic fields. As for the toric
code in a field, one can also show a duality mapping between the
so-called topological color code (TCC) \cite{bombin1_distilation} on
the dual honeycomb lattice, a quantum-spin model relevant for
topological quantum computation, in a magnetic field and the BWTF
\cite{ssj}.

In contrast, the physics of the BWTF at finite temperatures is to the best of our knowledge unknown. The latter is interesting, since thermal and quantum fluctuations are present simultaneously and one may expect tricritical behavior between the first-order phase transition at zero temperature and the second-order phase transition at zero field. Consequently, a detailed study of the finite-temperature phase diagram of the BWTF is the major focus of this paper. To this end we perform large-scale stochastic series expansion (SSE) quantum Monte Carlo (QMC) simulations. Furthermore, we discuss the low-energy spectral properties as well as the zero-temperature phase transition of the BWTF by comparing high-order series expansions about the low- and high-field limit using perturbative continuous university transformations (pCUTs) \cite{pcut3,pcut4} with the numerical data obtained by QMC. Here the pCUT allows the set up of a quasi-particle picture for the elementary excitations of the BWTF \cite{ssj,ssj2}.

The outline of the paper is as follows: In Sec.~\ref{model}, we introduce the Baxter-Wu model on the triangular lattice. The BWTF model is obtained by adding a transverse magnetic field in the $x$-direction, and we make the connection to the TCC in a parallel field on the dual honeycomb lattice. Thereafter, Sec.~\ref{methods} contains a detailed discussion on the methods used in this work. In Sec.\ref{pcut}, we review the basic concepts of 
the pCUT method and we prepare the reader for Sec.~\ref{finite-lattice} where we briefly review the quantum finite-lattice method combined with pCUTs to study the 
low-energy physics of the system on finite triangular clusters up to high orders in perturbation. The series expansion results are further given in Sec.~\ref{SEResult}, while Sec.~\ref{sec:QMCmain} presents the specificities of the QMC algorithm. Our results for the BWTF both at zero and finite temperature are presented in the main section of this paper (Sec.~\ref{phase_transition}), while the conclusion Sec.~\ref{conclusion} summarizes our main findings.

\section{Model and mapping}
\label{model}
\begin{figure}
\centerline{\includegraphics[width=6cm]{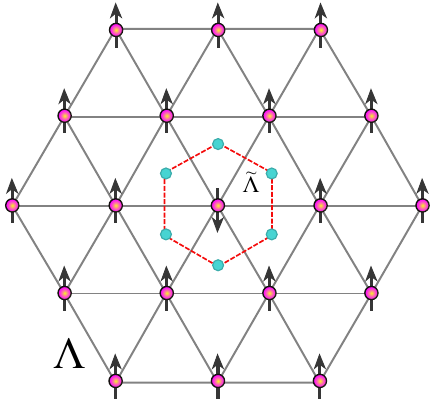}}
\caption{(Color online) Illustration of the triangular lattice $\Lambda$. Quantum spins 1/2 are located at the vertices of the lattice. Additionally, we sketch the elementary spin flip excitation above the ferromagnetic ground state of the BW-model with unit cell $\ket{\uparrow\uparrow\uparrow}$. This excitation can be created by the single operator $\sigma_i^x$ acting on site $i$ which flips the eigenvalues of the six face operators sharing this site. Alternatively, the spin flip excitation can also be seen as six excitations centered on triangles which form the dual honeycomb lattice $\widetilde{\Lambda}$.}
\label{triangular_latt}
\end{figure}

In this section, we introduce the BWTF. Additionally, we show a duality between the BWTF and the TCC in a parallel field which has been already discussed in Ref.\onlinecite{ssj}.

Let us consider a triangular lattice denoted by $\Lambda$ as illustrated in Fig.~\ref{triangular_latt}. Quantum spins 1/2 are located at the vertices of the lattice. The three spins on each triangle interact with each other by a three-spin interaction of strength $J>0$ such that the Baxter-Wu model is given by\cite{baxter2}
\be
H_{\rm BW}=-J\sum_{\langle ijk\rangle} \sigma^z_i\sigma^z_j\sigma^z_k , 
\label{H_BW}
\ee
where $\sigma^z$ is the ordinary $z$-Pauli matrix and the sum is taken over all triangles of the lattice. 

Each three-spin interaction (which we also call a face operator) in Eq.~(\ref{H_BW}) has eigenvalues $\pm 1$. Ground states of the Baxter-Wu model correspond therefore to states where all these eigenvalues of the face operators are $+1$. On a triangular lattice, the ground state of the system is 4-fold degenerate. The degenerate states can be represented by $\ket{\uparrow\uparrow\uparrow}$, $\ket{\uparrow\downarrow\downarrow}$, $\ket{\downarrow\uparrow\downarrow}$ and, $\ket{\downarrow\downarrow\uparrow}$ which corresponds to the configuration of the spins on the three sublattices. Here $\uparrow$ ($\downarrow$) denotes the +1 (-1) eigenvalue of the $\sigma^z$ Pauli operator. The ground-state energy of the system on a lattice with $N_f$ triangular faces is then given by $E_0=-N_fJ$. Furthermore, excited states are denoted by the total number of $-1$ eigenvalues of the face operators each of which costs the energy $2J$. Consequently, the first-excited states are $4N_f$-fold degenerate and have the total 
energy $E_1=E_0+2J$. The system is therefore gapped with an equidistant energy spectrum.

Using a duality transformation \cite{ssj}, one can introduce pseudospin-$\frac{1}{2}$ operators $\tau^z$ living on the dual honeycomb lattice of the triangular faces. The energetic properties of the Baxter-Wu model are then just given by an effective magnetic field
\be
H^{\rm dual}_{\rm BW}=-J\sum_{i \in \widetilde{\Lambda}} \tau^z_i \quad .
\label{H_BW_Map}
\ee
Here the sum runs over the vertices of the dual honeycomb lattice $\widetilde{\Lambda}$ which corresponds to the centers of triangles in the dual Baxter-Wu model. Let us stress that this duality is only valid for the spectrum, but any information on degeneracies is clearly lost. Indeed, the effective magnetic field has a unique polarized ground state with pseudo-spins pointing in the field direction and elementary excitations are dressed spin flips.

Our major objective is to add quantum fluctuations to the classical Baxter-Wu model on the triangular lattice by introducing a transverse magnetic field. We therefore consider the following Hamiltonian:
\be
H_{\rm BWTF}=-J\sum_{\langle ijk \rangle} \sigma^z_i\sigma^z_j\sigma^z_k -h \sum_i \sigma^x_i ,
\label{H_BWTF}
\ee
where the first (second) sum runs over the faces (sites) of the triangular lattice. In the following we focus on $h>0$.

A single $\sigma^x_i$ flips the eigenvalues of the six face operators sharing the site $i$ on the triangular lattice as illustrated in Fig.~\ref{triangular_latt}. In the dual pseudo-spin representation, the action of the transverse magnetic field is equivalent to flipping the 
six pseudo-spins of the corresponding hexagon. The BWTF after the duality mapping then reads
\be
H^{\rm dual}_{\rm BWTF}=-J\sum_{i \in \widetilde{\Lambda}} \tau^z_i-h \sum_{p\in \widetilde{\Lambda}} \chi_p\quad ,
\label{H_BWTF_Map}
\ee
where $p$ denotes a hexagon of the dual lattice $\widetilde{\Lambda}$ and \mbox{$\chi_p=\tau^x_1\tau^x_2\tau^x_3\tau^x_4\tau^x_5\tau^x_6$} are plaquette operators where the numbers ($1\ldots 6$) correspond to the six sites of hexagon $p$ (see also Fig.~\ref{triangular_latt}). 

The just derived dual Hamiltonian is closely related to the TCC in a single parallel field on the honeycomb lattice \cite{ssj}. The TCC is a topological stabilizer code consisting of two types of plaquette operators which all commute pairwise with each other such that eigenvalues $\pm 1$ of all these operators are conserved quantities \cite{bombin1_distilation}. One of the two types of plaquette operators is identical to $\chi_p$ introduced above. The other type of plaquette operator commutes with a magnetic field in the $z$-direction. As a consequence, the low-energy physics of the TCC in a single parallel field is contained in the sector where all of the remaining conserved plaquette operators are $+1$. Therefore, the BWTF is, up to a constant, isospectral to this sector of the TCC in a single parallel field \cite{ssj}.

\section{Methods}
\label{methods}

\subsection{Perturbative continuous unitary transformation}
\label{pcut}

Consider Hamiltonians of the form $H=H_0+xV$ where $H_0$ is the unperturbed part, $V$ the perturbation, and $x$ an expansion parameter, the pCUT method is applicable to those models that satisfy the following conditions \cite{pcut3}:
\begin{itemize}
\item The unperturbed part $H_0$ has an equidistant spectrum bounded from below. It can then be written as $H_0=E_0+Q$ where $E_0$ is the unperturbed ground-state energy and $Q$ counts the number of bare excitations.
\item The perturbation can be expressed as \mbox{$V =\sum_{n=-N_{\rm max}}^{N_{\rm max}} T_n$} where $T_n$ increments 
(decrements, if $n < 0$) the number of excitations (quasi-particles) by $n$ such that $\comutator{H_0}{T_n}=nT_n$. 
\end{itemize}
The pCUT method maps, order by order in $x$, the initial Hamiltonian $H$ to an effective one $H_{\rm eff}$ which conserves the number of quasi-particles (QP) \cite{pcut3,pcut4}
\be
H_{\rm eff}=E_0+Q+\sum_{k=1}^{\infty}x^{k} \sum_{\overline{m}=0 }C(m_1 \ldots m_k) T_{m_1}\ldots T_{m_k}\, ,
\label{effectiv_h}
\ee
such that $Q$ commutes with the effective Hamiltonian $\comutator{H_{\rm eff}}{Q}=0$. The first sum runs over the order of perturbation, $k$, while the second one runs over all possible permutations of 
$\{m_1,m_2,\ldots,m_k\}$ with \mbox{$m_i \in \{-N_{\rm max}, \ldots, N_{\rm max}\}$} which satisfy the condition \mbox{$\overline{m}=\sum_i m_i=0$}.
The coefficients $C(m_1 \ldots m_k)$ can be computed model independently as exact rational numbers up to high orders in perturbation\cite{pcut3}. 

Physically, operator sequences $T_{m_1}\ldots T_{m_k}$ in order $k$ represent quantum fluctuations of approximately a length scale $k$. These quantum fluctuations can be calculated in the thermodynamic limit by exploiting the linked-cluster theorem \cite{ser-exp,pcut3,pcut4}. The latter implies that only processes which take place on (finite) connected clusters contribute to matrix elements of $H_{\rm eff}$. Consequently, the pCUT method provides exact results up to the computed perturbative order directly in the thermodynamic limit.

\subsubsection{High-field limit ($h\gg J$)}
In this subsection, we apply the pCUT method to the BWTF about the high-field limit. We therefore rewrite Eq.~(\ref{H_BWTF}) 
\be
\frac{H_{\rm BWTF}}{2h}=-\frac{1}{2}\sum_i \sigma^x_i-\frac{J}{2h}\sum_{<ijk>} \sigma^z_i\sigma^z_j\sigma^z_k , 
\label{H_hf}
\ee
such that the magnetic field represents the unperturbed part and the Baxter-Wu model is the perturbation. The ground state of the pure field term is a polarized phase in the $x$-direction and the system has an equidistant spectrum. Therefore, both necessary conditions for pCUTs are fulfilled. Elementary excitations are local spin flips on the sites of the triangular lattice. They can be conveniently described in terms of hard-core bosons represented by creation and annihilation operators $b^{\phantom{\dagger}}_i$ and $b^{\dagger}_i$. One finds
\bea
\frac{H}{2h}&=& -\frac{N}{2}+\sum_i n_i+\frac{J}{2h} \sum_{<i,j,k>} (b^{\dagger}_i b^{\dagger}_j b^{\dagger}_k +b^{\dagger}_i b^{\dagger}_j b^{\phantom{\dagger}}_k \nonumber \\
&& +b^{\dagger}_i b^{\phantom{\dagger}}_j b^{\dagger}_k +b^{\phantom{\dagger}}_i b^{\dagger}_j b^{\dagger}_k +{\rm h.c.})\nonumber \\
&=& E_0+Q+\frac{J}{2h} (T_{3}+T_{1}+T_{-1}+T_{-3}),
\label{H_hf_Tn}
\eea
where $N$ is the number of triangular lattice sites and $n_i=b^{\dagger}_i b^{\phantom{\dagger}}_i$ is the local density on site $i$. 

The three-spin interactions of the Baxter-Wu model do therefore change the number of QPs by $n=\{\pm1,\pm3 \}$.
The pCUT then maps Hamiltonian (\ref{H_hf_Tn}) to an effective Hamiltonian $H^{\rm hf}_{\rm eff}$ which conserves the number of QPs. 
This allows the investigation of the low-energy properties of the BWTF in the high-field limit along the lines discussed in Ref.~\onlinecite{ssj}. 
The series expansion results for the ground-state energy per site $\varepsilon_0^{\rm hf}$ as well as the 1-QP gap $\Delta^{\rm hf}$ are presented in Sec.~\ref{SEResult}.

\subsubsection{Low-field limit ($J\gg h$)}
Here we aim at a linked cluster expansion for the low-field limit of the BWTF. To this end we can benefit from the duality mapping discussed in Sec.~\ref{model} and we can apply the pCUT method to the dual Hamiltonian (\ref{H_BWTF_Map}) on the dual honeycomb lattice \cite{ssj}. Since the first term of Eq.~(\ref{H_BWTF_Map}) is an effective field term, the unperturbed part is exactly the same as the one in the high-field limit and we can again apply the pCUT method using spin-flip excitations of the dual model as the appropriate QPs. The plaquette operators $\chi_p$ represents the perturbation which are six-spin interactions on hexagons. The latter operators change the number of QPs by \mbox{$n=\{0,\pm2,\pm4 ,\pm6\}$}. Consequently, the dual Hamiltonian (\ref{H_BWTF_Map}) can be recast into the following form
\be
\frac{H^{\rm dual}_{\rm BWTF}}{2J}= E_0+Q+\frac{h}{2J} \sum_{n=\{0,\pm2,\pm4 ,\pm6\}} T_n,
\label{H_lf_Tn}
\ee
where $E_0=-N$. The explicit expressions of $T_n$ operators have been already given in Ref.~\onlinecite{ssj}.

The pCUT then maps Eq.~(\ref{H_lf_Tn}) to an effective Hamiltonian $H^{\rm lf}_{\rm eff}$ in the low-field limit. The energetics of the BWTF in the low-field limit is then accessed by calculating the ground-state energy per site and the low-energy gap of the QP excitations.

However, one should note that although a single spin flip on one site of the dual honeycomb lattice is the lowest excited state, the 1-QP sub-block is not the most relevant sector close to the phase transition \cite{ssj}. In the original BWTF, the transverse field $\sigma^x_i$ creates (out of the unperturbed ground state) six QPs on the six neighboring triangles which share site $i$. This condition implies that this 6-QP excitation having all six spins of a hexagon flipped to be the most relevant low-energy excitation. We therefore discuss $\varepsilon_0$ and the corresponding 6-QP gap of the BWTF in the low-field limit in the next section.

\subsection{Quantum finite-lattice method}
\label{finite-lattice} 

As we have already outlined in Sec.~\ref{pcut}, the pCUT method maps the initial problem to a QP conserving Hamiltonian which is a sum over virtual fluctuations that are represented by sequences of the $T_n$ operators. This property has important consequences when the method is applied to Hamiltonians with local $T_{n}$ operators
\cite{pcut3}. By local, we mean $T_n=\sum_{\vec{i}} T_{n,\vec{i}}$ where $\vec{i}$ are a finite number of neighboring sites.

In Ref.~\onlinecite{qflm}, it has been shown that $H_{\rm eff}$ can alternatively be represented by an infinite sum of nested commutators of these local operators. This property is a direct consequence of the so-called linked-cluster theorem \cite{ser-exp,pcut3,Gelfand} which states that only those processes contribute to the matrix elements of the effective Hamiltonian which take place on linked clusters.

According to Gelfand {\it et al.}\cite{Gelfand}, the series expansion of any extensive quantity $F$ per site on a lattice $\Lambda$ can be expressed as a sum over linked clusters $c$:
\be
F(\Lambda)/N=\sum_c L(\Lambda,c) W(c),
\label{Fc}
\ee
where
\be
W(c)=F(c)-\sum_{c'\subset c} W(c')\quad.
\label{Wc}
\ee
Here $N$ is the number of lattice sites, $L(\Lambda,c)$ is the number of embeddings of cluster $c$ per lattice site, and $W(c)$ is the reduced weight of the cluster $c$.
In order to obtain $W(c)$, one first calculates the series expansion of $F$ on cluster $c$ and then subtracts the contributions of all subclusters $c'\neq c$ which can be
embedded in $c$ ($c'\subset c$). In the present work, $F$ denotes the matrix elements of $H_{\rm eff}$ which correspond to the ground-state energy or hopping amplitudes of elementary quasi-particles.

The finite-lattice method (FLM) was first introduced by Enting {\it et al.}~\cite{flm-square} in the framework of classical statistical mechanics and applied to the Ising model
on the square lattice. The main idea of the FLM was to consider rectangular clusters $C_{m\times n}$ with $m\times n$ sites and their embeddings in the lattice $\Lambda$.
Then any extensive quantity such as the free energy is calculated on rectangular clusters and the physical quantity in the thermodynmic limit is obtained by proper 
summation and subtraction rules. The main benefit of this method is that the total number of clusters is dramatically decreased compared to a full graph expansion. 
Furthermore, the embedding number can be determined analytically for some lattices such as the square lattice \cite{flm-square}. The FLM was first brought to the realm of 
quantum many-body problems by Dusuel {\it et al.}~in Ref.~\onlinecite{qflm} where it has been applied to the transverse field Ising model and the XXZ model on the square lattice.
  
We have applied Enting's finite-lattice method for the triangular lattice\cite{flm-triangle1,flm-triangle2} to the BWTF model and we calculated the ground-state energy 
and the 1-QP gap of the system in the high-field limit. In contrast to the square lattice, there is no algebraic relation for the embedding factor of subclusters for 
the triangular lattice. One therefore has to determine these factors numerically. The interested reader can find an efficient algorithm for generating the subclusters 
and obtaining their embedding number in Ref.~\onlinecite{flm-triangle2}. After identifying the clusters and their embedding number, we calculate the matrix elements of $H_{\rm eff}$ by acting on each cluster and thereafter subtracting the contributions of subclusters $c'\subset c$ using Eqs.~(\ref{Fc} and \ref{Wc}). 

\subsection{Series expansion results}
\label{SEResult}
\subsubsection{High-field results}
\label{SEResult:hf}
Using the quantum finite-lattice method on the triangular lattice, we calculated the ground-state energy $e_0^{\rm hf}$ and the 1-QP gap $\Delta^{\rm hf}$ of the system about the high-field limit for $h=1$:
\bea
e_0^{\rm hf} &=&-1-\frac{1}{3}J^2-\frac{19}{216} J^4-\frac{5359}{34020} J^6 \\ \nonumber
&& -\frac{500690327}{1371686400} J^8-\frac{74305313819}{72013536000}J^{10} \quad,
\eea 
\bea
\Delta^{\rm hf} &=& 2-24 J^2+64 J^4-\frac{268712}{81} J^6 \\ \nonumber
&& + \frac{37389778504}{893025} J^8-\frac{29786981411535707}{20253807000} J^{10}\label{gap-small} \quad .
\eea
Note that both series obtained by the quantum finize-lattice method correspond exactly to the ones calculated in Ref.~\onlinecite{ssj} for the TCC in a parallel field.

\subsubsection{Low-field results}
\label{SEResult:lf}
Finally, we also give the series expansion results of the ground-state energy $e_0^{\rm lf}$ and the 6-QP gap $\Delta^{\rm lf}$ (relevant mode close to the transition) of the system in the low-field limit for $J=1$ obtained by pCUTs:
\bea
e_0^{\rm lf} &=& -2-\frac{1}{12}h^2-\frac{1}{864}h^4-\frac{19}{155520} h^6 \\ \nonumber
&& -\frac{1133}{238878720} h^8-\frac{12026279}{27088846848000} h^{10} \quad,
\eea
\bea
\Delta^{\rm lf} &=& 12-\frac{22}{3}h^2+\frac{88}{27} h^4-\frac{413}{72} h^6\\ \nonumber
&& +\frac{20157041}{1749600} h^8-\frac{1446718370831}{52907904000} h^{10}\label{gap-large} \quad .
\eea
The series are obtained by acting with $H_{\rm eff}$ on the appropriate hexagonal clusters in different orders of perturbation as explained in Ref.~\onlinecite{ssj}. 

\subsection{Quantum Monte Carlo}
\label{sec:QMCmain}
\subsubsection{Algorithm}
\label{sec:QMC}
Working in the standard basis where the Ising interaction is diagonal,
all non-vanishing off-diagonal elements of the BWTF are negative for $h>0$: it is therefore amenable to QMC simulations with no sign problem. However, it is non-trivial to
devise an efficient algorithm due to the specific form of the three-body Ising interaction. The SSE QMC technique with efficient directed loop updates~\cite{SSE} is indeed usually formulated for two-body interactions. For specific models however, one can construct non-local loop algorithms in a slightly different fashion as first formulated by Sandvik~\cite{Sandvik2003} for the TFIM model in the basis where the Ising interaction is diagonal. We refer the reader to Ref.~\onlinecite{Sandvik2003} for full details of the algorithm, and just state the key point: for the TFIM, the matrix
element of an Ising (SSE) vertex is unchanged when both spins (all SSE vertex legs) are flipped. Unfortunately, this property does no longer hold for the BWTF due to the odd number of sites in the Ising interaction. However, we can use a
similar idea: before starting a global loop update in the SSE configuration, one can freeze randomly one of the three
sublattices A, B or C~\cite{remark3} by not allowing the loop to touch any of the spins in this sublattice. Then, we use exactly the same rules as for the TFIM case~\cite{Sandvik2003}, and flip all legs of SSE vertices \emph{except} those on the frozen sublattice,
i.e. four legs among six. By doing so, the weight of the configuration is unchanged and the acceptance rate is
one. In the limit of vanishing transverse field, this algorithm is identical to the cluster algorithm first proposed by Evertz and Novotny~\cite{EvertzNovotny} for the classical Baxter-Wu model, exactly as the algorithm proposed by Sandvik~\cite{Sandvik2003} reduces to the Swendsen-Wang~\cite{SwendsenWang} algorithm for the Ising model.

This remark makes us believe that our QMC algorithm for the quantum Baxter-Wu model is slightly less efficient than the TFIM algorithm, as it was already shown~\cite{EvertzNovotny} for the classical limit $h\rightarrow 0$ that the dynamical critical exponent (characterizing the algorithm efficiency to decorrelate Monte-Carlo samples)  was larger than for the Swendsen-Wang algorithm. We nevertheless manage to simulate close to criticality large samples of size $N=L \times L$ up to $L=128$ at finite temperature, and up to $L=15$ in the ground-state. We have checked by comparing with exact diagonalizations (ED) on small lattices that our QMC implementation does reproduce all quantities within error bars.

In the QMC simulations, we measure several observables which can be grouped in two types: observables related to the order parameter of the ordered magnetic phase of the BWTF model, and observables related to energy. The sections below give their definitions and expected scaling close to a phase transition.

\subsubsection{Observables based on order parameter}

To define properly the order parameter, let us first discuss the symmetries of the model Eq.(\ref{H_BWTF}). It is clearly invariant when one performs a $\pi$ rotation in spin space (i.e. spin inversion) for spins sitting on any two among the three sublattices A, B or C (this fact is explicitly used in the construction of the QMC algorithm, see Sec.~\ref{sec:QMC}).  As a consequence, any sublattice magnetization $$m_\alpha=1/(N/3)\sum_{i\in \alpha} \sigma_i^z$$ 
(where $\alpha=$A, B or C) will have a vanishing expectation value $\langle m_\alpha \rangle=0$ on finite systems.
This is also true for the expectation value of any cross-correlation $\langle m_A m_B \rangle = \langle m_B m_C \rangle =\langle m_A m_C \rangle=0$. 

Using the above symmetry (which is recovered in our simulations within error bars), we can construct an estimator of the square of the order parameter 
\begin{equation}
m_s^2 = \frac{m_A^2 + m_B^2 + m_C^2}{3}
\end{equation}
which expectation value reaches its maximum value in the four classical ground states. The same quantity was also measured in classical simulations~\cite{Shchur2010}. Close to a continuous phase transition at finite temperature, we expect the following scaling form~\cite{PrivmanBook}:
$$\langle m_s^2 \rangle = L^{-2\beta / \nu} f((T-T_c)\, L^{1/\nu})$$
where $\beta$ is the order parameter critical exponent, $\nu$ is the correlation length critical exponent, $T_c$ is the critical temperature and $f$ is a universal scaling function. The exponents of the classical BW model are those of the $4$-states Potts model in two dimensions (with no logarithmic correction): $\beta=1/12$, $\nu=2/3$.  For a first-order transition, we expect a jump at the critical point.

A useful way of localizing a continuous phase transition is to consider the associated Binder cumulant
\begin{equation}
U_L = 1 - \frac{3}{5}\frac{\langle m_s^4 \rangle}{\langle m_s^2\rangle^2}.
\end{equation}
Normalization is such that $U\rightarrow 2/5=0.4$ in the ordered phase, and $U\rightarrow 0$ in the disordered phase.  For a continuous phase transition, the following scaling form is expected:
$$U_L =g((T-T_c)\, L^{1/\nu})$$
with $g$ being a universal scaling function, such that for various sizes $L$, Binder cumulants should cross at the critical point. The exponent $\nu$ can furthermore be extracted from a scaling plot. For first-order phase transitions one does not observe unique crossings and the Binder cumulant reaches very large negative values just above the critical point (in the disordered phase), which should scale as $L^d=L^2$. This has been explained phenomenologically by Vollmayr {\it et al.}~\cite{Vollmayr1993} and can be simply understood as resulting from the double-peak structure of the order parameter distribution. Some continuous phase transitions may also display negative values for  Binder cumulants, but the minimum does {\it not} scale as $L^d$.

Observables based on the $z$ components of spins are easily measured in QMC as they are diagonal in the chosen $\sigma^z$ basis. 

\subsubsection{Observables based on energy}
\label{sec:obsE}
Singularities in energy and its derivative can clearly signal phase transitions both at finite and zero temperature.
For finite temperature, the specific heat per site:
$$C_v/N=\frac{\langle H^2 \rangle - \langle H \rangle^2}{NT^2}$$
is well-known to display a singularity at a finite-$T$ phase transition. For a continuous transition, we expect the following scaling form: $$C_v/N = L^{\alpha/\nu} h((T-T_c)\, L^{1/\nu})$$ where $\alpha$ is the specific heat critical exponent and $h$ is another universal scaling function. For the BW model, we know that $\alpha=2/3$, in agreement with hyperscaling relation $\alpha=2-d\nu$. 
For a first-order phase transition, we rather expect a (volume-scaling) divergence $C_v/N \sim L^2$.

At zero temperature, the analog of the specific heat is the second-derivative of the ground-state energy, 
$$\chi_E=-\frac{1}{N}\frac{\partial^2 E_0}{\partial h^2}$$ 
with respect to the field strength. For a second-order phase transition, this quantity should scale at the transition as~\cite{Albuquerque2010}: $$ \chi_E \sim L^{(2/\nu)-(d+z)}$$ where $z$ is the dynamical exponent ($d=2$ here). For a first-order phase transition on the other hand, we expect a volume divergence: $\chi_E \sim L^{d}$.

Another interesting insight on the phase transition is provided by  the energy Binder cumulant\cite{Challa1986}~:
$$V_L=1-\frac{\langle H^4\rangle}{3\langle H^2\rangle^2}.$$
This quantity is equal to $2/3$ away from the phase transition (both in the disordered and ordered phase)\cite{Challa1986}. For a  continuous transition, the same limit $V(T_c)=2/3$ is obtained in the thermodynamic limit. On the other hand, a finite dip (with $V=V^* \neq 2/3$) is observed in the thermodynamic limit at a first order phase transition point\cite{Challa1986}, and can be directly related to the two different characteristic energies of each phase. For finite systems, the value of the dip is expected\cite{Challa1986} to scale as $V_{\rm min}(L)=V^*+A/L^d$. For a continuous transition, a dip can also be found on finite systems, but it should vanish in the thermodynamic limit where $V(T_c)=2/3$ is recovered. We are not aware of the precise derivation of a finite-size scaling form of the energy Binder cumulant at a continuous transition, although general arguments~\cite{Challa1986,Martinos2005} suggest the following scaling $V_{\rm min}(L)-2/3\sim L^{-d+\alpha/\nu}$.

Even though they correspond to off-diagonal observables, moments of the Hamiltonian are easily computed within the QMC method (by computing the appropriate moment of the expansion order~\cite{SSE}). The second-derivative of the ground-state energy is also directly accessible as a simple response function within the SSE technique~\cite{SSE,Albuquerque2010}.

\section{Numerical results for phase transitions}
\label{phase_transition}
\subsection{Quantum phase transition}
\label{qpt}

In this section, we investigate the zero-temperature phase transition
driven by the magnetic field, combining pCUT and QMC results. We have
calculated with pCUTs the ground-state energy per site, $e_0$, as well
as the 1-QP gap, $\Delta=\omega({\bf K}=0)$, of the system for the
small- and high-field limit of the BWTF model (see Sec.~\ref{SEResult} for the full
expressions). Note that exactly the same series have been obtained in
Ref.~\onlinecite{ssj} when studying the TCC in a single parallel
field. Nevertheless it is in our opinion valuable to give them again
here in the appropriate units for the BWTF. 

\begin{figure}
\includegraphics[width=\columnwidth]{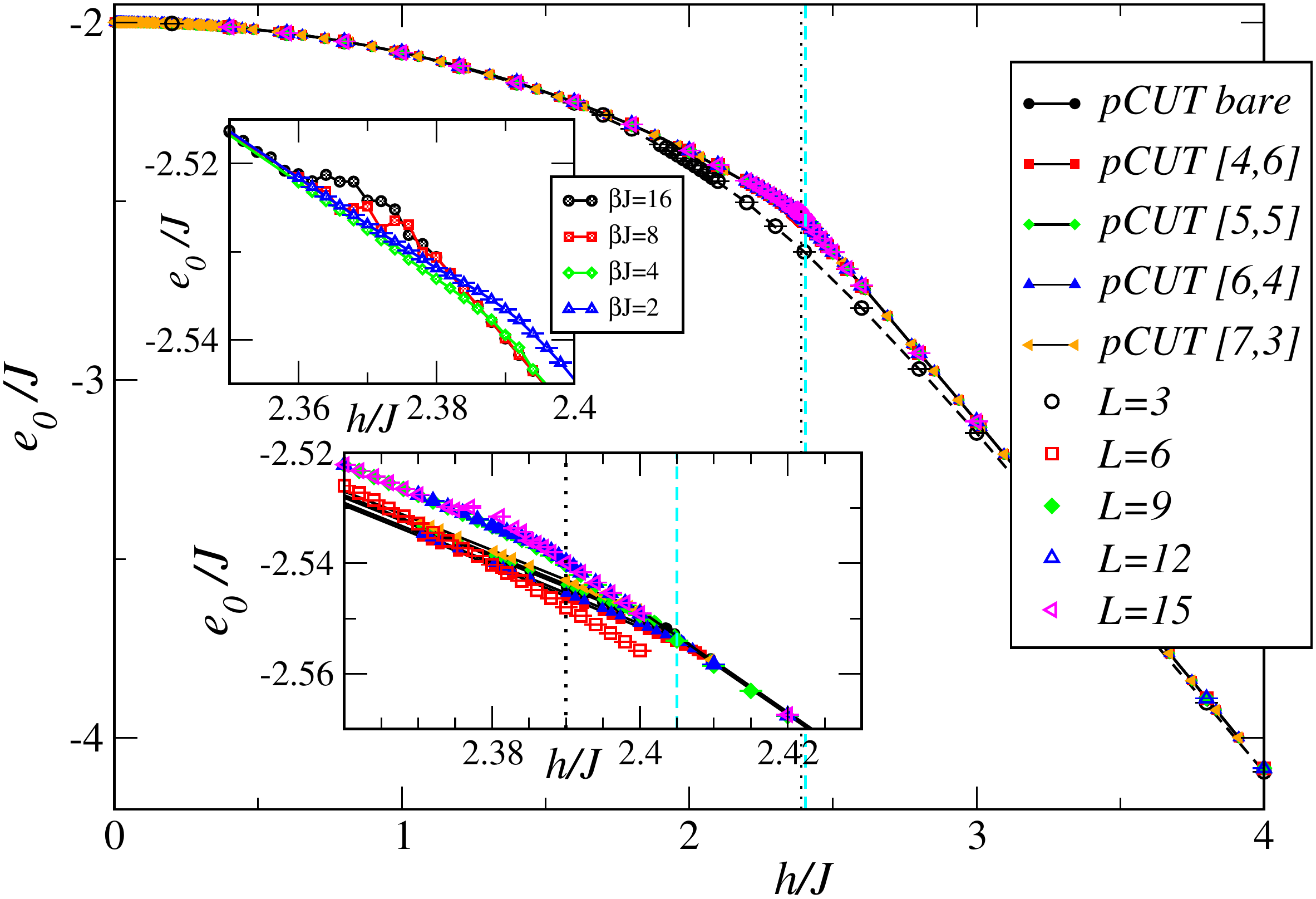}
\caption{(Color online) Ground-state energy per site, $e_0$, of the BWTF model as a function of the transverse field $h$ obtained by QMC (open symbols) and pCUT (filled symbols). QMC data is obtained for fixed inverse temperature $\beta J=16$ and for different linear sizes $L$. For $L=3$, a dashed line indicate the $T=0$ exact result. The vertical dotted lines (respectively the vertical dashed lines) correspond to the location of the first-order phase transition $h/J \simeq 2.39$ ($h/J \simeq 2.405$) deduced from the different QMC observables (from the crossing point of the two high-order series expansions \cite{ssj}). Bottom inset: Zoom on the region close to the phase transition. Top inset: Equilibration effects for the energy curve of a $L=18$ QMC sample, as a function of inverse temperature $\beta$. In the latter case, we do not plot error bars when simulations are not converged.}
\label{fig:e0}
\end{figure}

\begin{figure}
\includegraphics[width=\columnwidth]{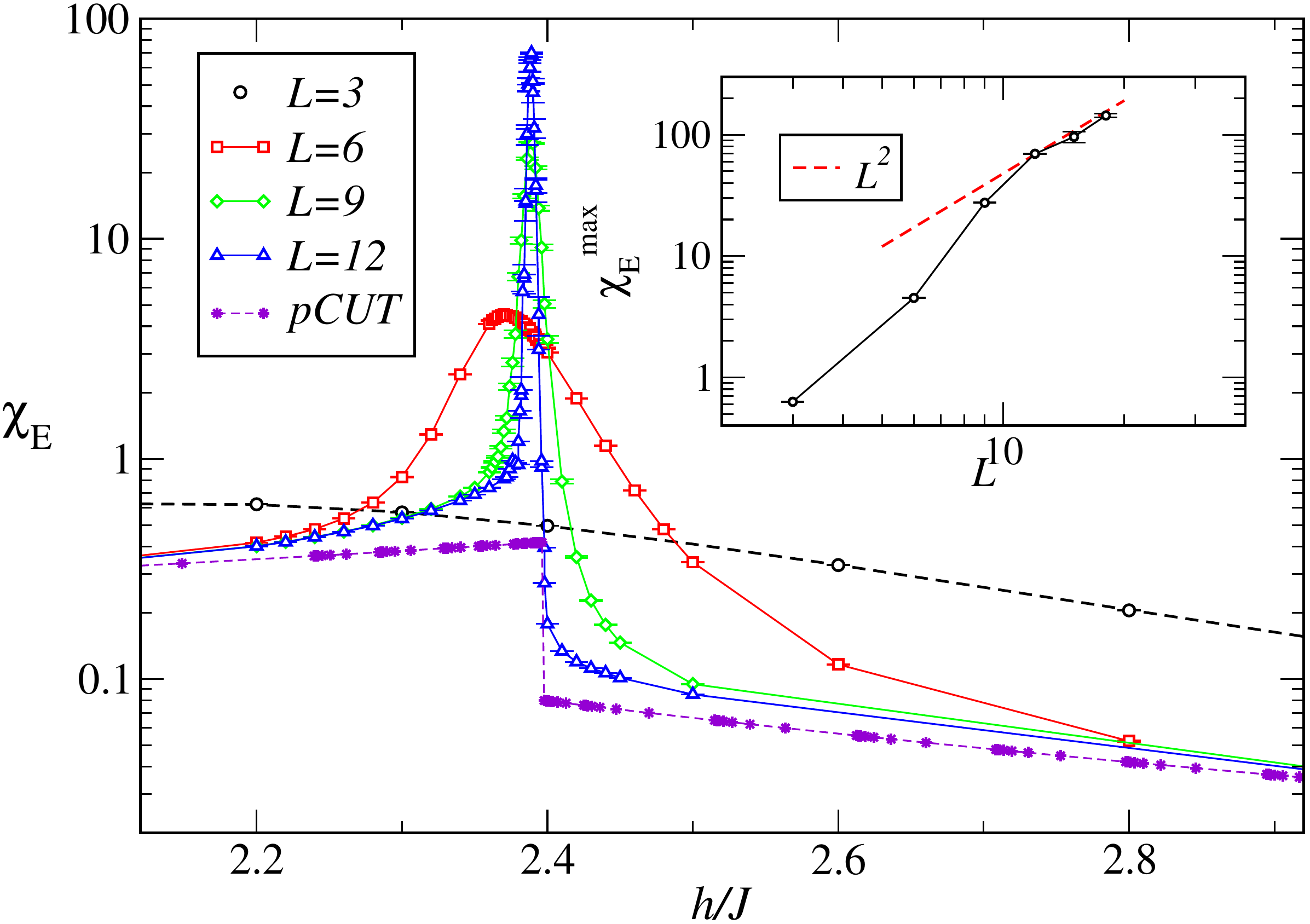}
\caption{(Color online) $\chi_E=-\partial^2 e_0/\partial h^2$ vs transverse field $h$ at low-temperature $\beta J=16$ for various sizes. Inset: maximum of this quantity vs $L$ showing a divergence (the dotted line displays the expected $L^2$ divergence for comparison).}
\label{fig:chiE_vs_h}
\end{figure}

Combining the low- and high-field series expansions, we display $e_0/J$ as a 
function of $h/J$ in Fig.~\ref{fig:e0}. The crossing point of both 
expansions signals a first-order phase transition close to $h\simeq 2.405J$
 in full agreement with Ref.~\onlinecite{ssj}. We refer the interested 
reader to the latter reference for a detailed discussion on capturing the
first-order transition point by analyzing the 1-QP energy gap. Additionally, we also show the 
QMC expectation value for different system sizes, at a fixed inverse temperature $\beta J
=16$. As we can clearly see from the zoom close to $h\simeq 2.4J$ in
the bottom inset, there is a hint of a discontinuity in the ground-state
energy curve which also pleads in favor of a first-order phase
transition. We provide below a more refined analysis using the second-derivative of the energy with respect to the field, namely $\chi_E$ (see Fig.~\ref{fig:chiE_vs_h}).

However, we observe that the QMC energy curves for 
larger sizes $L=18$ and $L=24$ deviate from the collapse (data not shown). 
We attribute this to a lack of equilibration at this temperature of our QMC
simulations for these large system sizes. This is confirmed by considering
the single sample $L=18$ for different values of inverse temperature
(see top inset of Fig.~\ref{fig:e0}), where the average energy at lower
temperature is incorrectly {\it larger} than for higher temperature in
this field region. Note that in such cases, a careful statistical analysis does confirm that simulations are \emph{not} converged, so we do not show error bars. Such hysteretic effects in Monte Carlo simulations
are typical of a {\it first-order} phase transition.

\begin{figure}
\includegraphics[width=\columnwidth]{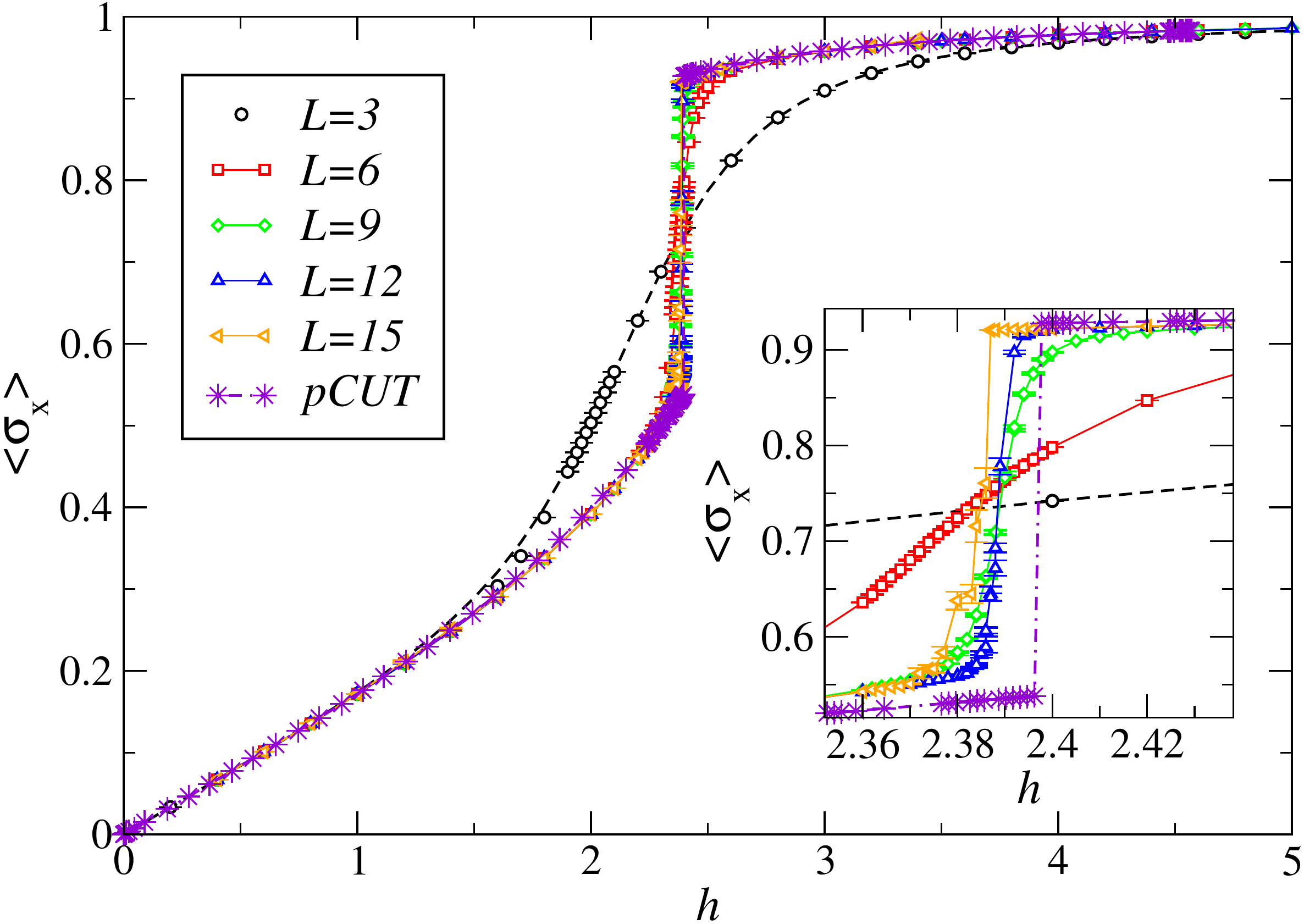}
\caption{(Color online) (a) Magnetization per site along field $\langle \sigma^x \rangle$ vs transverse field $h$ at low-temperature $\beta J=16$ for various sizes obtained by QMC and pCUT results. Inset: zoom on the transition region where a sharp jump is forming. Note that lines are guide to the eyes except for $L=3$ which displays the exact $T=0$ results from ED.}
\label{fig:Sx_vs_h}
\end{figure}

\begin{figure}
\includegraphics[width=\columnwidth]{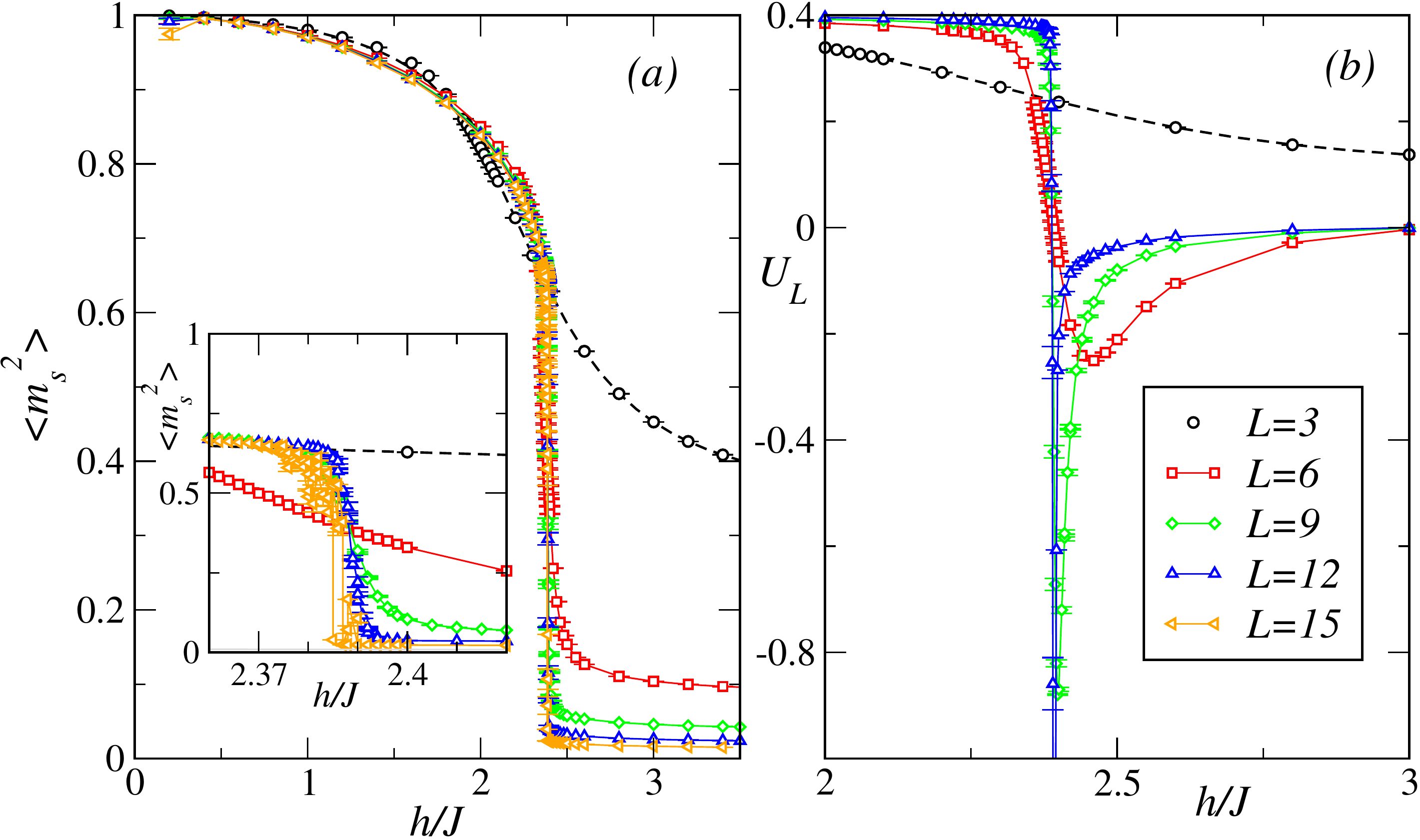}
\caption{(Color online) (a) Sublattice magnetization density square $\langle m_s^2 \rangle$ vs transverse field $h$ at low-temperature $\beta J=16$ for various sizes. Inset: zoom on the transition region where a sharp jump is forming. (b) :  Binder cumulant $U_L$ vs transverse field $h$ at low-temperature $\beta J=16$ for various sizes. Note that lines are guide to the eyes except for $L=3$ which displays the exact $T=0$ results from ED.}
\label{fig:ms2_binder_vs_h}
\end{figure}

This is confirmed by the following results for other QMC observables
(we now fix the inverse temperature to be $\beta J=16$, and limit the
system size to $L\leq 15$ to ensure correct convergence). 
In Fig.~\ref{fig:Sx_vs_h}, we plot the magnetization per site along the field $\langle \sigma^x \rangle$ versus transverse field $h$. While this is not an order parameter, this quantity displays a rather nice behavior with a paramagnetic response at small field, a sharp jump close to $h/J \simeq 2.39$, and then approaches saturation. Note that the pCUT fully agrees with the QMC results displaying a jump at slightly larger values $h/J \simeq 2.395$ as shown in Fig.~\ref{fig:Sx_vs_h}.
The (square of the) order parameter (Fig.~\ref{fig:ms2_binder_vs_h}a) similarly 
displays a marked jump versus field at the same critical value of the field
$h\simeq 2.39 J$ (see inset). The order parameter Binder cumulant $U_L$
also displays a jump (see Fig.~\ref{fig:ms2_binder_vs_h}a), as well as
a large negative (apparently diverging with $L$) value above the
transition field.  As seen later, this may not indicate a first-order
transition though.  In order to confirm the discontinuous nature of
the transition, we display in Fig.~\ref{fig:chiE_vs_h} the behavior
versus field of the second-derivative of the ground-state energy
$\chi_E$, which clearly diverges strongly at the transition. The
divergence appears (see inset) to approximatively scale as $L^2$ (and
maybe even with a larger exponent, but this is probably due to the
limited range of $L$ available at this low temperature). Note however 
that the divergence expected in $\chi_E$, and seen in QMC data, 
comes from the non-analyticity of $e_0(h)$ and thus cannot be captured 
by the perturbative pCUT approach. 

Overall, this set of results clearly point towards a first-order quantum phase transition in the BWTF model
Eq.~(\ref{H_BW}). The QMC simulations point to \mbox{$h_c(T=0)\simeq 2.39J$} while the pCUT results tend 
to a slightly larger value \mbox{$h_c(T=0)\simeq 2.4J$}. This small difference might arise due to the convergence 
problem in QMC as explained above or due to the uncertainties in the pCUT originating from the extrapolation
of the series. 

\subsection{Finite temperature phase diagram}

We now turn to finite-temperature properties of the BWTF model. At $h=0$, the BW model is known to have a second-order transition in the 4-state Potts model universality class (without logarithmic corrections), i.e. with critical exponents $\alpha=\nu=2/3$, $\beta=1/12$, at the self-dual point $T_c= 2J/\log(\sqrt{2}+1) \simeq 2.269 J$. When switching on a small field $h$, it is not clear a priori  whether the transition will remain of second-order type or change to first-order, as will be the case eventually when $h \rightarrow h_c$ at zero temperature. 

Our QMC simulations discussed below indicate that the finite-temperature transition remains continuous in the same universality class as for the classical BW model in a large region of transverse field, up to at least \mbox{$h= 2.25 J$}. Additionally, we have indications that the transition is first-order for $h= 2.35 J$. In the following, we present a selected set of results for three values of the field: two in the continuous regimes ($h=0$ corresponding to the classical case, and $h=2J$) and one in the first-order regime ($h=2.35J$), which will allow us to contrast the different behaviors of several observables. We consider the $h=0$ classical limit explicitly as this allows to benchmark the methodology against exact results. The resulting full $(h,T)$ phase diagram is constructed and discussed in the Sec.~\ref{conclud}. 

\begin{figure}
\includegraphics[width=\columnwidth]{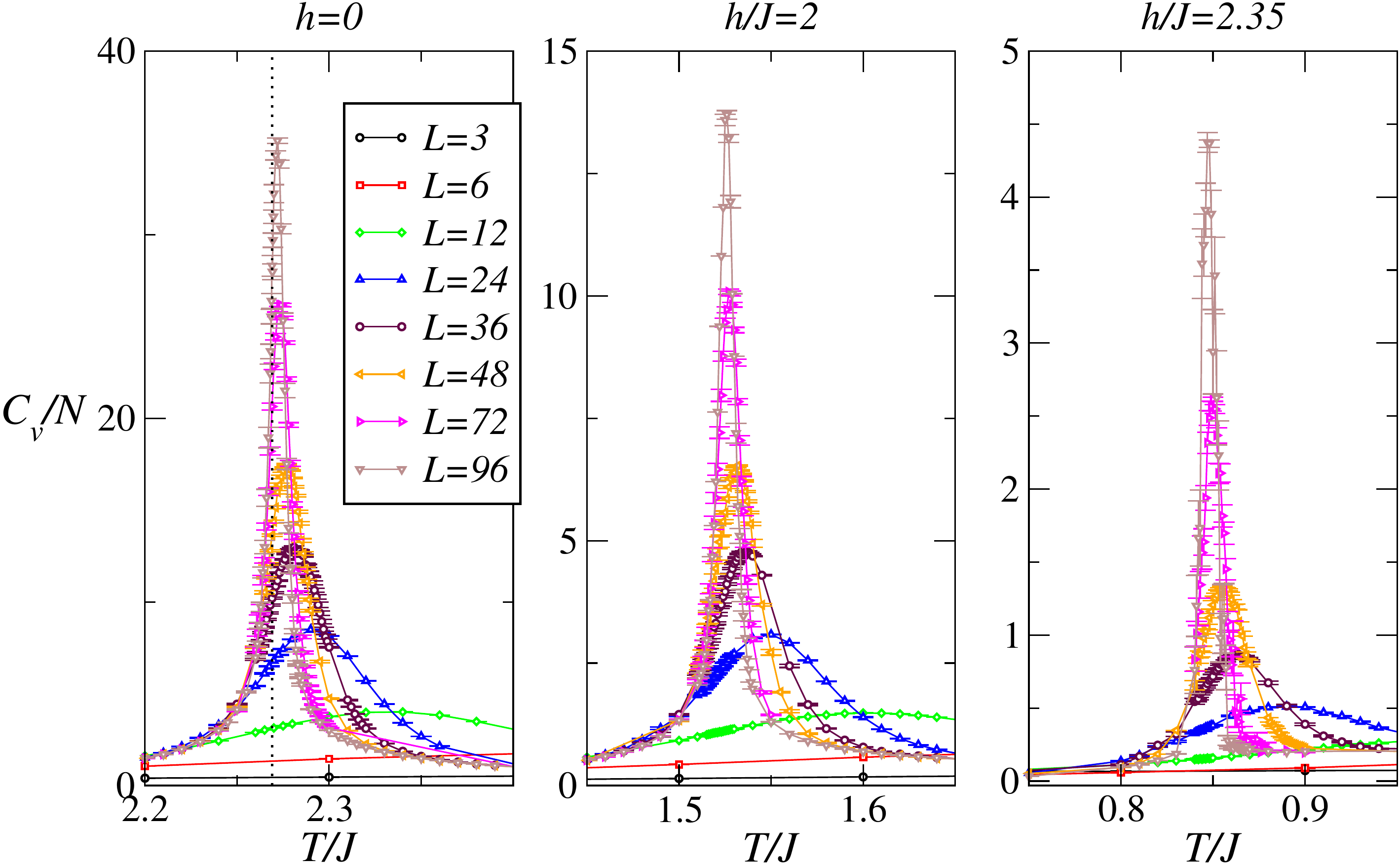}
\caption{(Color online) Specific heat $C_v$ vs temperature $T$ for various sizes. The three panels correspond to $h=0$, $h=2J$ and $h=2.35 J$ respectively. For $h=0$ (classical case), the exact $T_c$ is indicated by the dotted vertical line.}
\label{fig:Cv_panel}
\end{figure}

\begin{figure}
\includegraphics[width=\columnwidth]{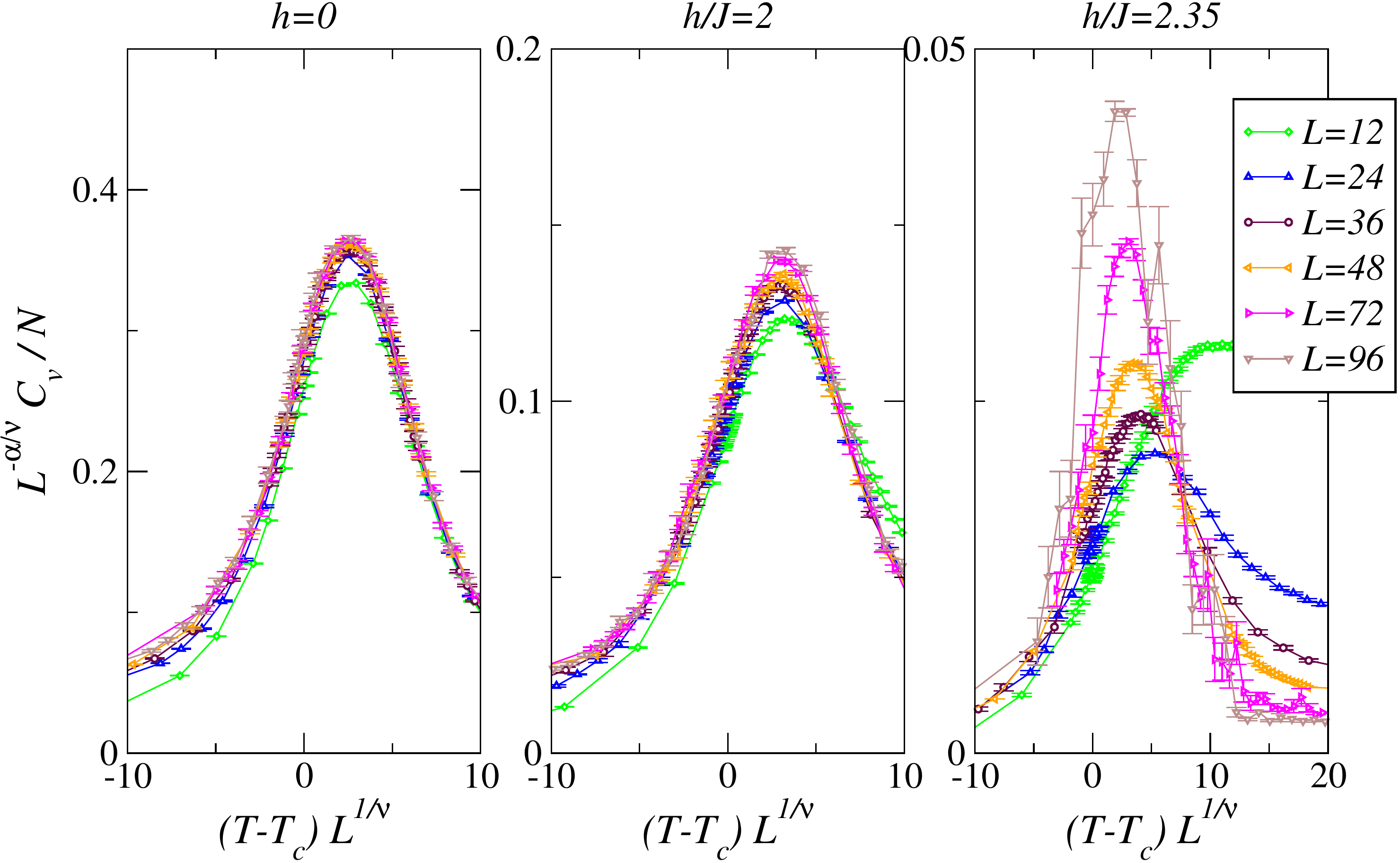}
\caption{(Color online) Collapse analysis using the same data as in Fig.~\ref{fig:Cv_panel}. Using the known results for BW model and the obtained $T_c$'s, we plot $C_v/L^{\alpha/\nu}$ vs $(T-T_c(h))\, L^{1/\nu}$ with $\alpha=\nu=2/3$.}
\label{fig:Cv_collapse_panel}
\end{figure}

\begin{figure}
\includegraphics[width=\columnwidth]{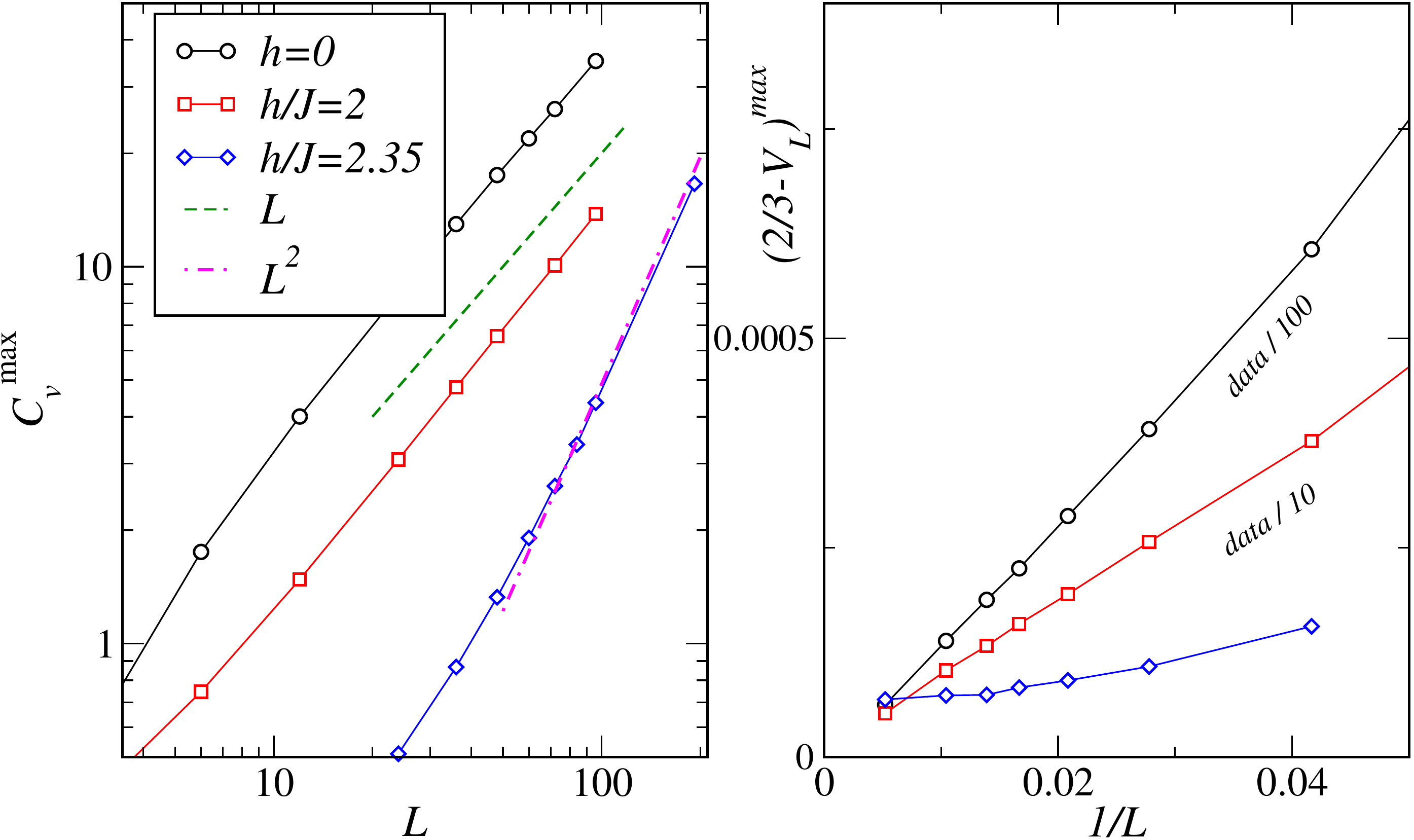}
\caption{(Color online) (a) Specific heat maximum vs length $L$ for various $h$. (b) Dip size of $V_L$ vs $1/L$ for various $h/J$ }
\label{fig:Cvmax_Vmin}
\end{figure}

We first consider energetics. Fig.~\ref{fig:Cv_panel} displays the specific heat $C_v/N$ as a function of temperature for the three selected values of the field. A  divergence is observed in all three cases, yet with a much more marked behavior for $h=2.35J$. The power-law envelope of the first two sets of curves suggests a continuous behavior, while the stronger divergence of $C_v/N$ at $h=2.35J$ rather indicate a first-order behavior. This is confirmed by the collapse plot shown in Fig.~\ref{fig:Cv_collapse_panel} where the data at $h=0$ and $h=2J$ can be well collapsed with the expected form at a continuous transition with the BW exponents. This is clearly not the case for the third panel of Fig.~\ref{fig:Cv_collapse_panel}. A confirmation is further obtained by considering the divergence of the maximum of $C_v/N$ as a function of system size (see Fig.~\ref{fig:Cvmax_Vmin}a): the linear behavior (corresponding to $\alpha / \nu=1$) for $h=0,2J$ contrasts with the (volume) scaling $L^2$ for $h=2.35J$, as expected for 
a first-order phase transition. The $T_c$'s obtained from the $C_v/N$ divergence are :  $T_c(h=0)=2.27J$ (in agreement with the exact result), $T_c(h=2J)=1.52J$ and $T_c(h=2.35J)=0.85J$.

\begin{figure}
\includegraphics[width=\columnwidth]{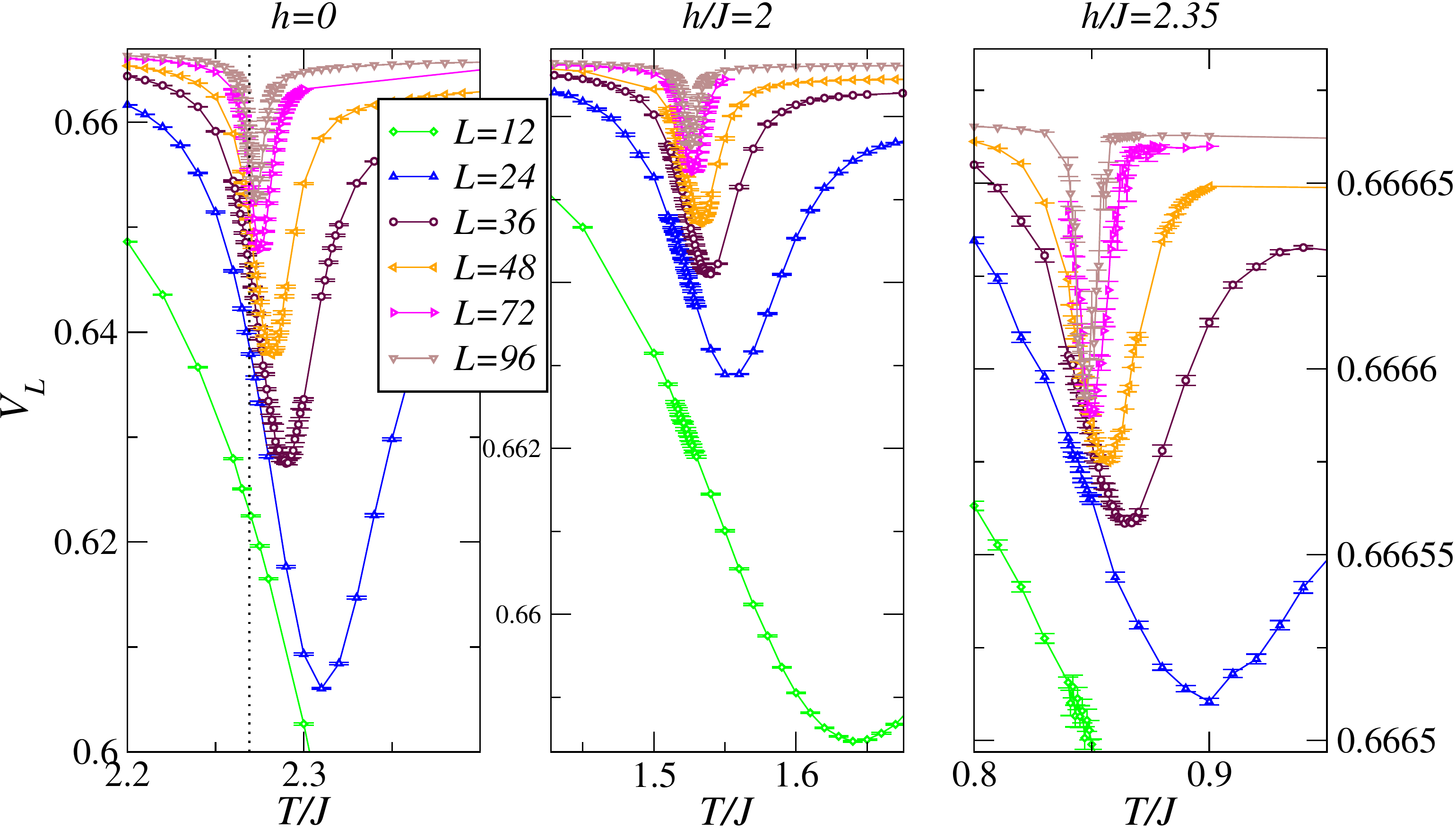}
\caption{(Color online) Energy Binder cumulant $V_L$ vs $T$ for various sizes and transverse fields. A dip in this quantity signals a phase transition (see text).}
\label{fig:V_panel}
\end{figure}

\begin{figure}
\includegraphics[width=\columnwidth]{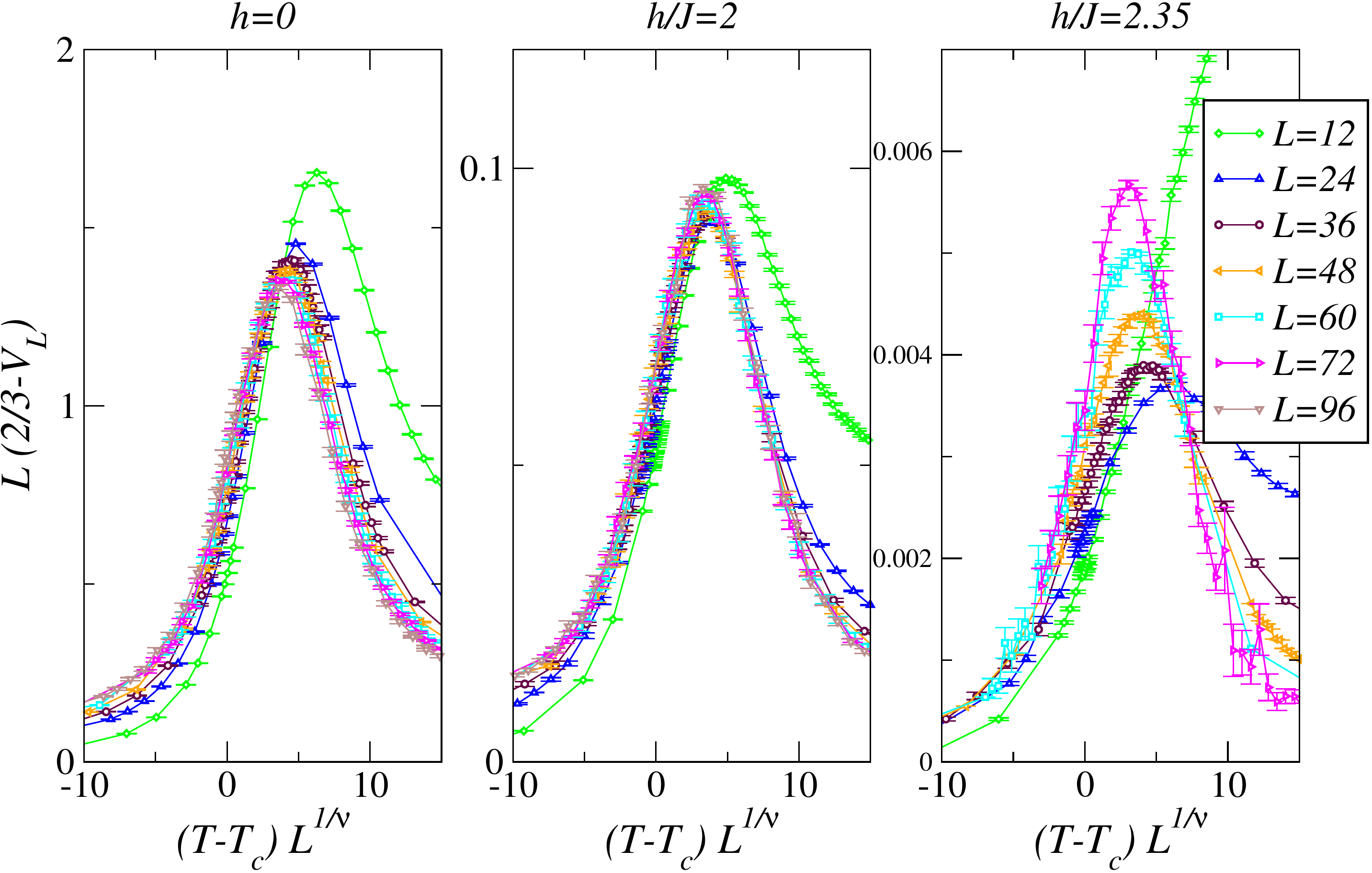}
\caption{(Color online) Tentative data collapse of the $V_L$ dip times $L$ vs $(T-T_c)L^{1/\nu}$ using the critical exponents of the BW model. A rather good collapse can be obtained for $h=0$ and $2J$, but not for $h=2.35J$.}
\label{fig:V_collapse_panel}
\end{figure}

An \emph{independent} check of the nature of the phase transitions is obtained by considering the energy Binder cumulant $V_L$. In all three cases, we observe a dip close to $T_c(h)$ (see Fig.~\ref{fig:V_panel}). However, the data at $h=0,2J$ have a different finite-size behavior as can be observed in Fig.~\ref{fig:Cvmax_Vmin}b: the minimum value $V_{\rm min}$ converges to $2/3$ with $1/L$ (as expected from the scaling ansatz presented in Sec.~\ref{sec:obsE} with $\alpha/\nu=1$), while $V_{\rm min}-2/3$ appears to reach a non-zero value with a different power-law $L^{-2}$ for $h=2.35J$. The existence and scaling of the dip in $V$ was also reported in $h=0$ classical studies~\cite{Martinos2005,Schreiber2005}. This is again a sign of a first-order phase transition for this latter field value, which is confirmed by the fact that a (phenomenological) collapse of the full $V_L$ curves can be obtained for $h=0,2J$, but not for $h=2.35J$ (see Fig.~\ref{fig:V_collapse_panel}).

\begin{figure}
\includegraphics[width=\columnwidth]{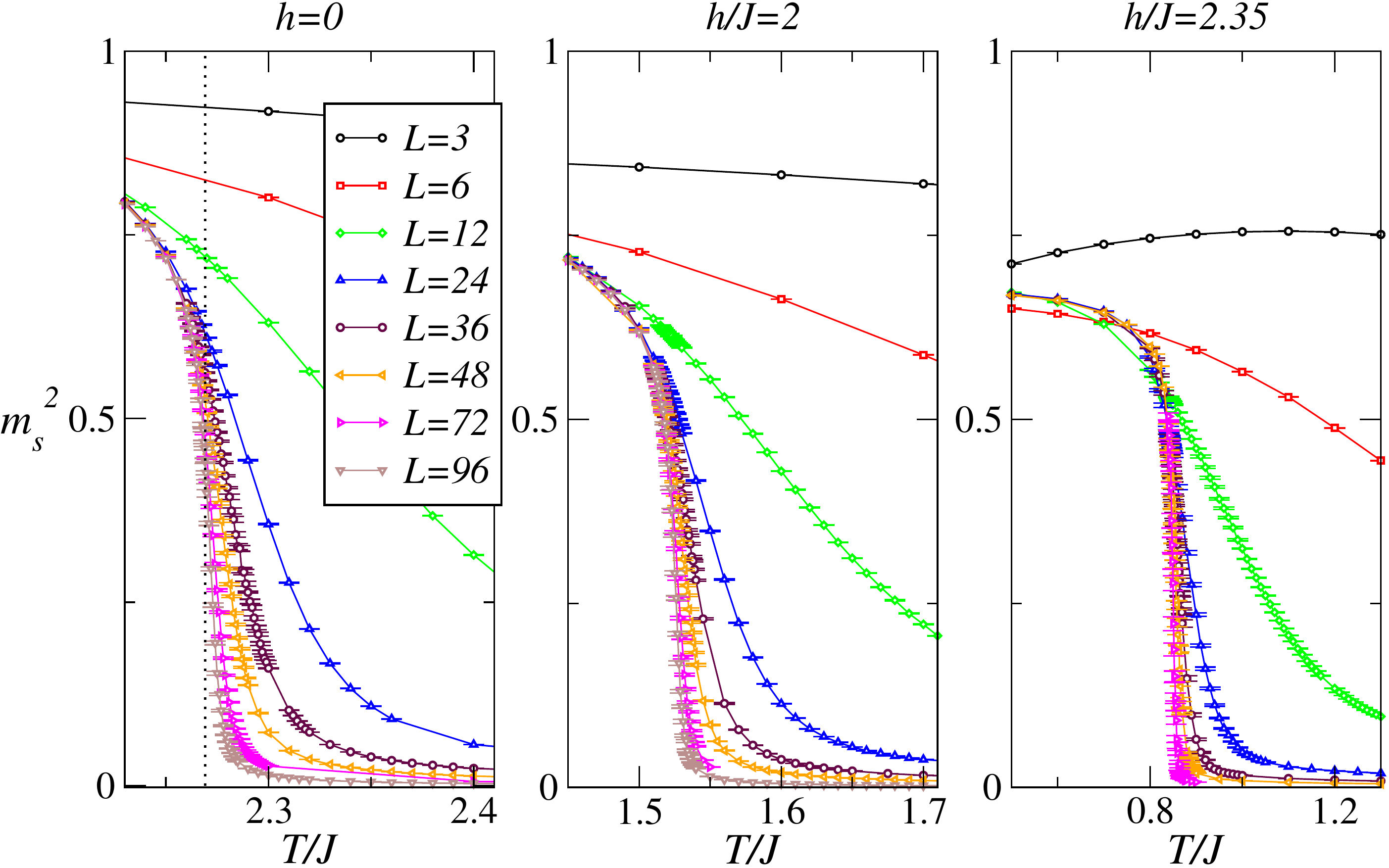}
\caption{(Color online) Order parameter squared $\langle m_s^2 \rangle$ as a function for temperature $T/J$ for different system sizes, for different field values.}
\label{fig:ms2_panel}
\end{figure}

\begin{figure}
\includegraphics[width=\columnwidth]{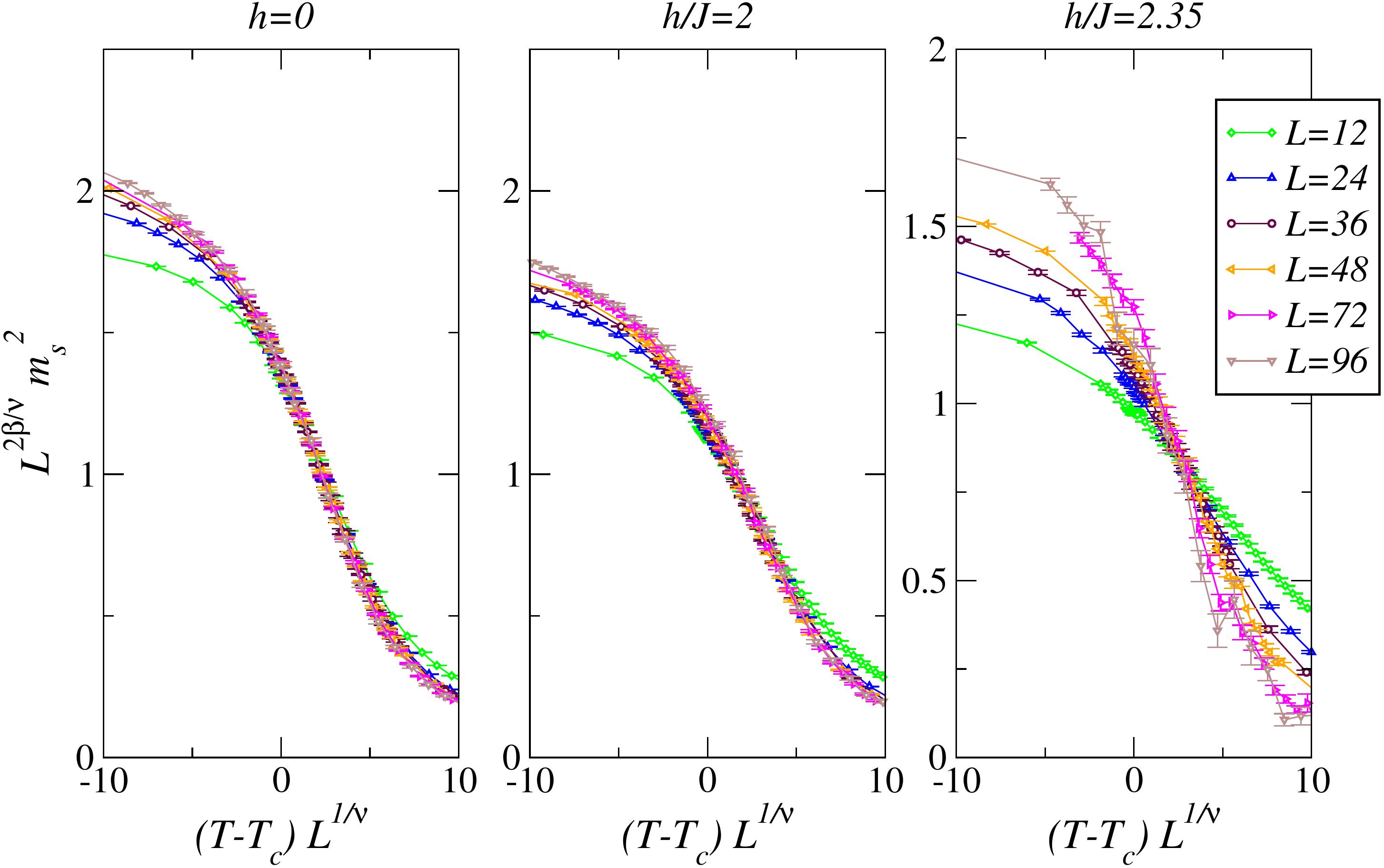}
\caption{(Color online) Scaling of the order parameter squared $\langle m_s^2 \rangle$ times $L^{2\beta/\nu}$ vs $(T-T_c)L^{1/\nu}$ with the values $\beta/\nu=1/8$ and $\nu=2/3$ showing an excellent data collapse (two first panels for $h=0$ and $h=2J$), and no collapse (right panel) for $h=2.35J$.}
\label{fig:ms2_collapse_panel}
\end{figure}

\begin{figure}
\includegraphics[width=\columnwidth]{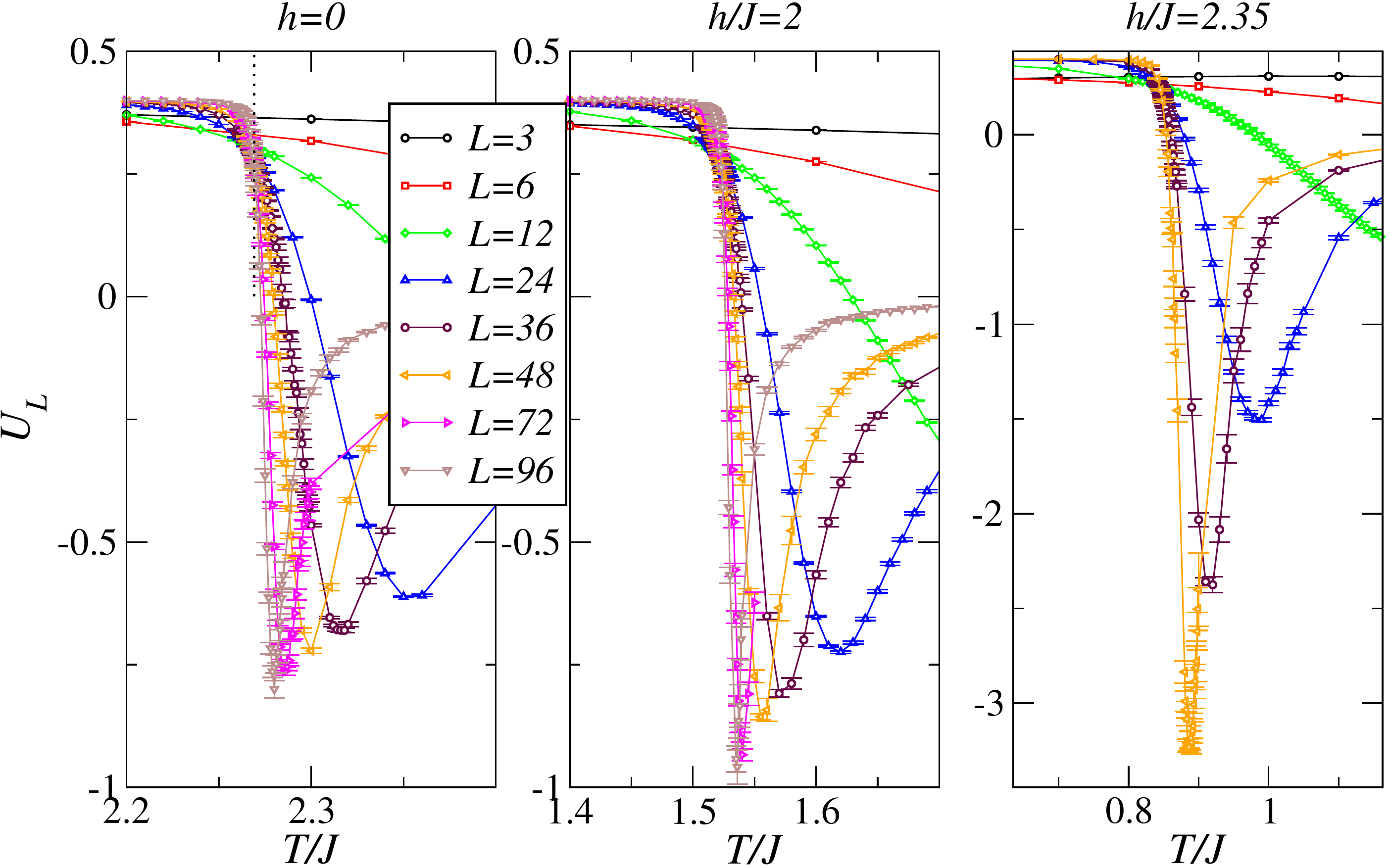}
\includegraphics[width=\columnwidth]{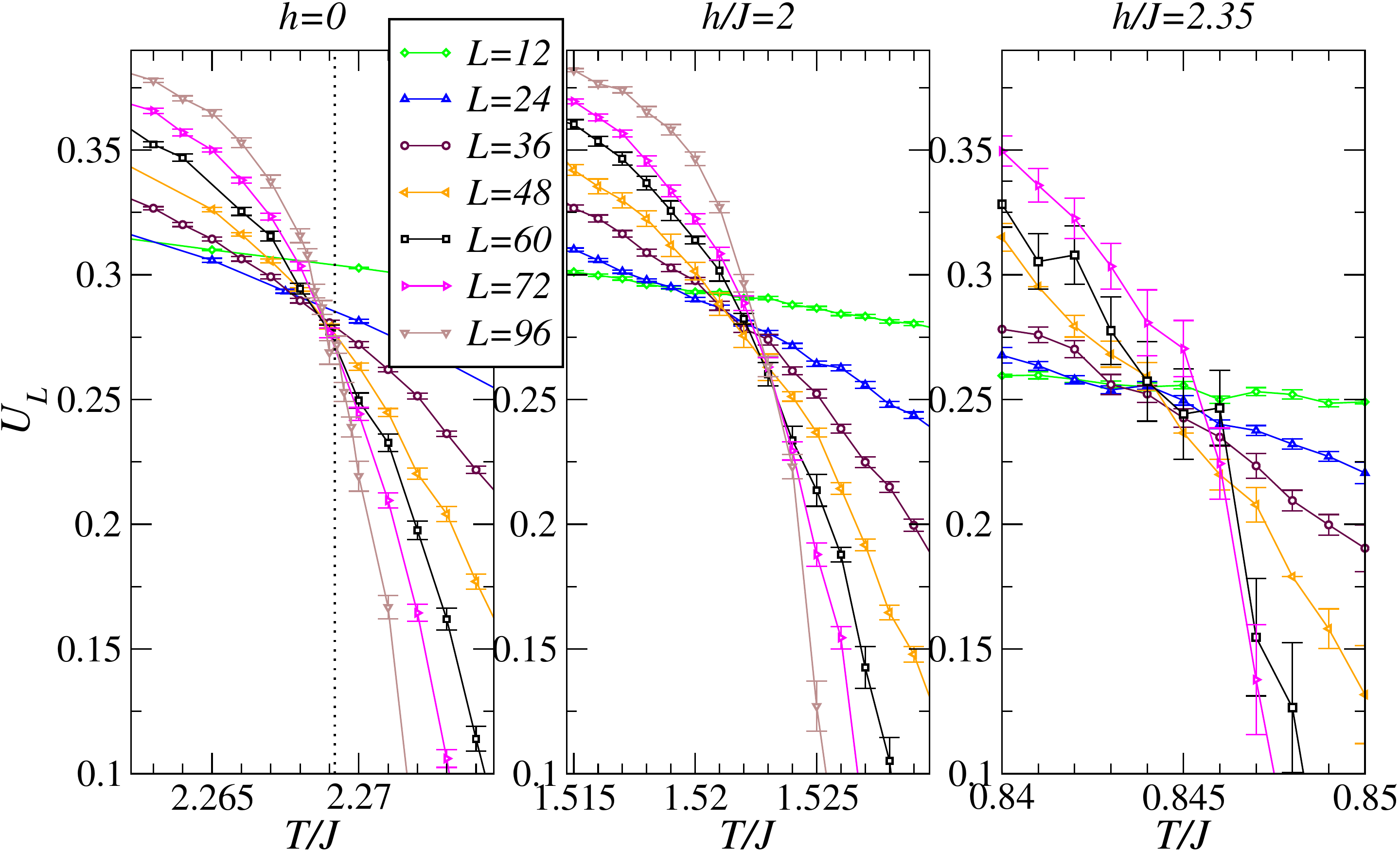}
\caption{(Color online) (a) Order parameter Binder cumulant $U_L$ vs $T$ for various sizes and transverse field $h$. (b) Zoom close to the transition temperature. For $h=0$, the exact $T_c$ is indicated as a dashed line. 
}
\label{fig:U_panel}
\end{figure}

\begin{figure}
\includegraphics[width=\columnwidth]{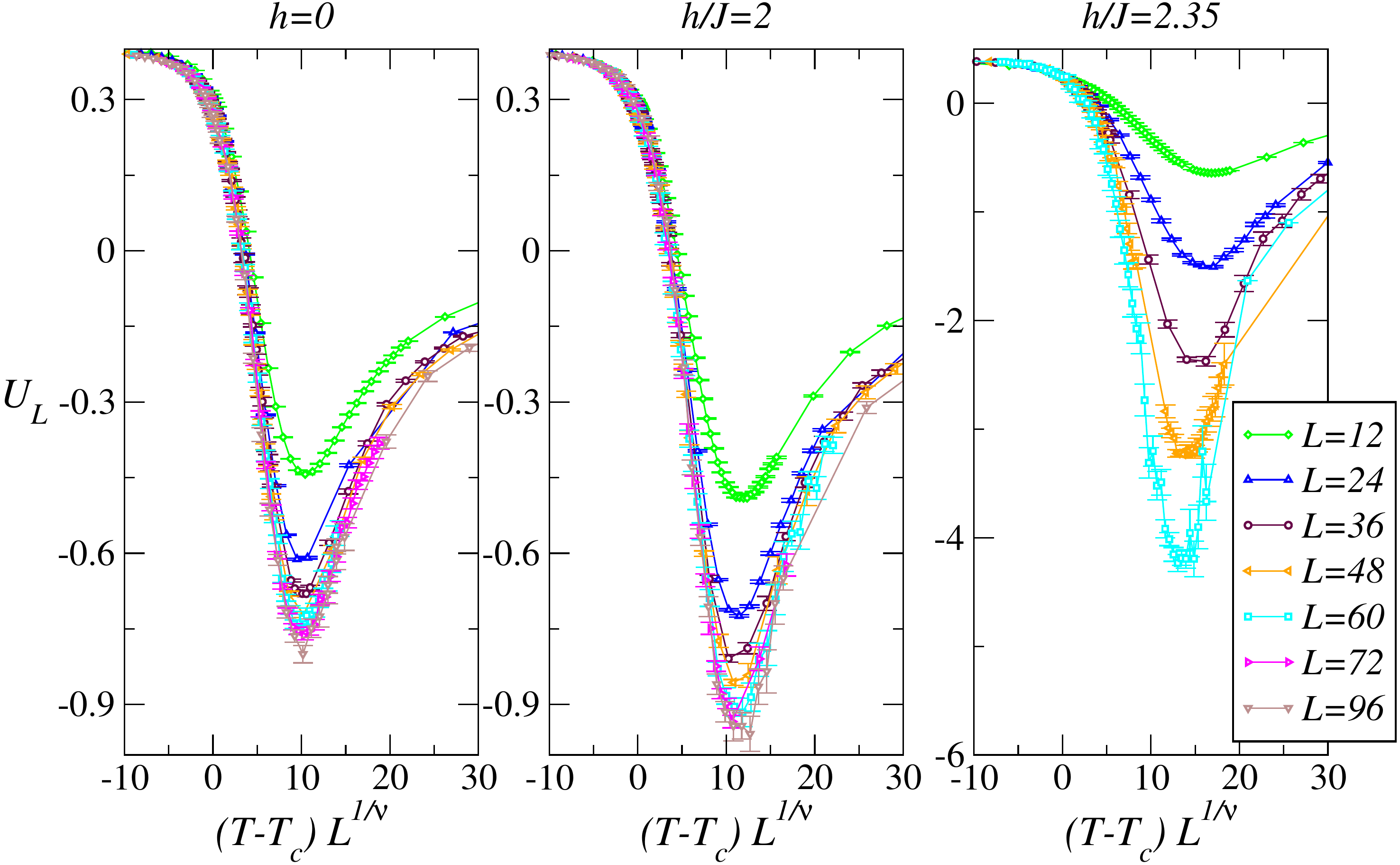}
\caption{(Color online) Scaling of the order parameter Binder cumulant $U_L$  vs $(T-T_c)L^{1/\nu}$ using the known exponents of the BW model.  }
\label{fig:U_collapse_panel}
\end{figure}

Fig.~\ref{fig:ms2_panel} represents the (squared) order parameter $m_s^2$ as a function of temperature, for the three selected field values and for different $L$. A rather sharp drop is observed at the values of $T_c$ estimated from the above energetics considerations. While the curves for different system sizes can be reasonably well collapsed with the BW exponents ($\beta / \nu=1/8$) for the field values $h=0,2J$ (see Fig.~\ref{fig:ms2_collapse_panel}), this is not the case for $h=2.35J$. The results for the magnetization Binder cumulant $U_L$ confirm this analysis, albeit with a further anomaly in the high-temperature phase. Indeed on the largest clusters, a nice crossing  for different $L$ is observed (with value close to 0.27) in Fig.~\ref{fig:U_panel} for $T_c (h=0)\simeq 2.269 J$ and $T_c (h=2J) \simeq 1.523 J$, but not for the largest field value where curves do not cross at a single point. We note also the existence of negative values for larger $T$ (for all fields): we find (data not shown) that 
this minimum does diverge in the thermodynamic limit for $h=2.35J$ but not for $h=0,2J$, therefore implying a first-order character for the largest field and a continuous nature for the two other field values. This is confirmed by the excellent data collapse for $U_L$ presented  in Fig.~\ref{fig:U_collapse_panel} obtained using the exact value $\nu=2/3$.

Note that we have also performed a systematic data collapse (without any prior knowledge) for all the above quantities. For instance, data analysis of the Binder cumulant $U_L$ for the three cases $h=0$, $h=2J$ and $h=2.35J$ respectively leads to estimates of $\nu=0.6701\pm 0.003$, $\nu=0.6664 \pm 0.004$ and $\nu =0.5205 \pm  0.01$. Such results are fully compatible with our claim that the first two cases are second-order phase transitions in the same universality class as BW model, while the third one is first-order with an effective $\nu=1/d=1/2$. 

\begin{figure}
\includegraphics[width=\columnwidth]{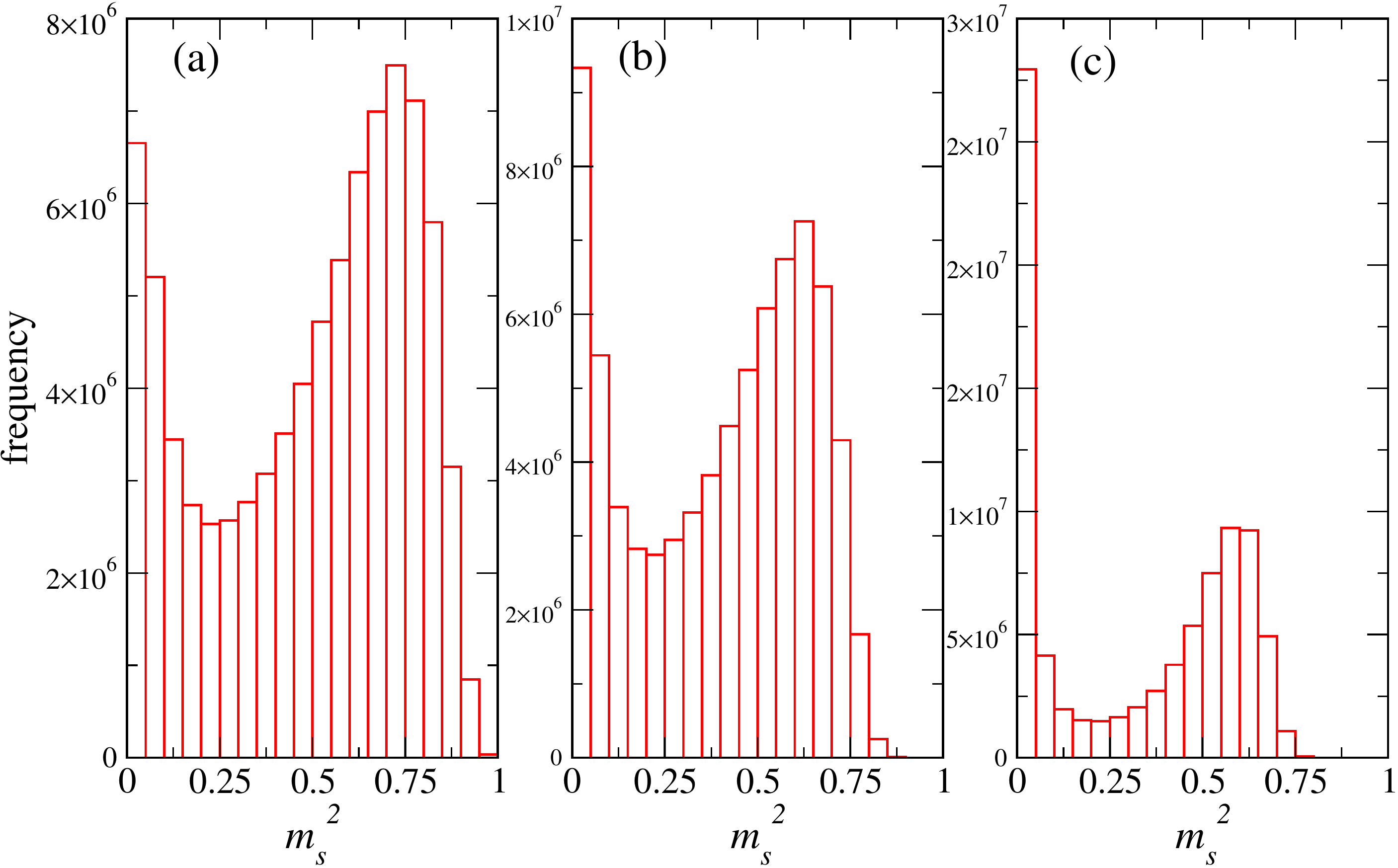}
\caption{(Color online) Order parameter histograms on $L=36$ cluster for (a) $h=0$, $T/J=2.274$; (b) $h/J=2$, $T/J=1.53$; (c) $h/J=2.35$, $T/J=0.86$. }
\label{fig:m2_histo}
\end{figure}

In order to provide a more physical picture in the change of nature of the phase transition when varying $h/J$, we provide now some analysis of the full energy and order parameter histograms. In Fig.~\ref{fig:m2_histo}, the order parameter histograms for various $h$ (taken at the transition, i.e. fixing $T/J$ at the maximum slope of $m_s^2(T)$) display some characteristic bimodal structure. This bimodal form is reinforced when increasing $h$, although there is no qualitative change, and is responsible for the negative $U_L$ values that were discussed previously. 

\begin{figure}
\includegraphics[width=\columnwidth]{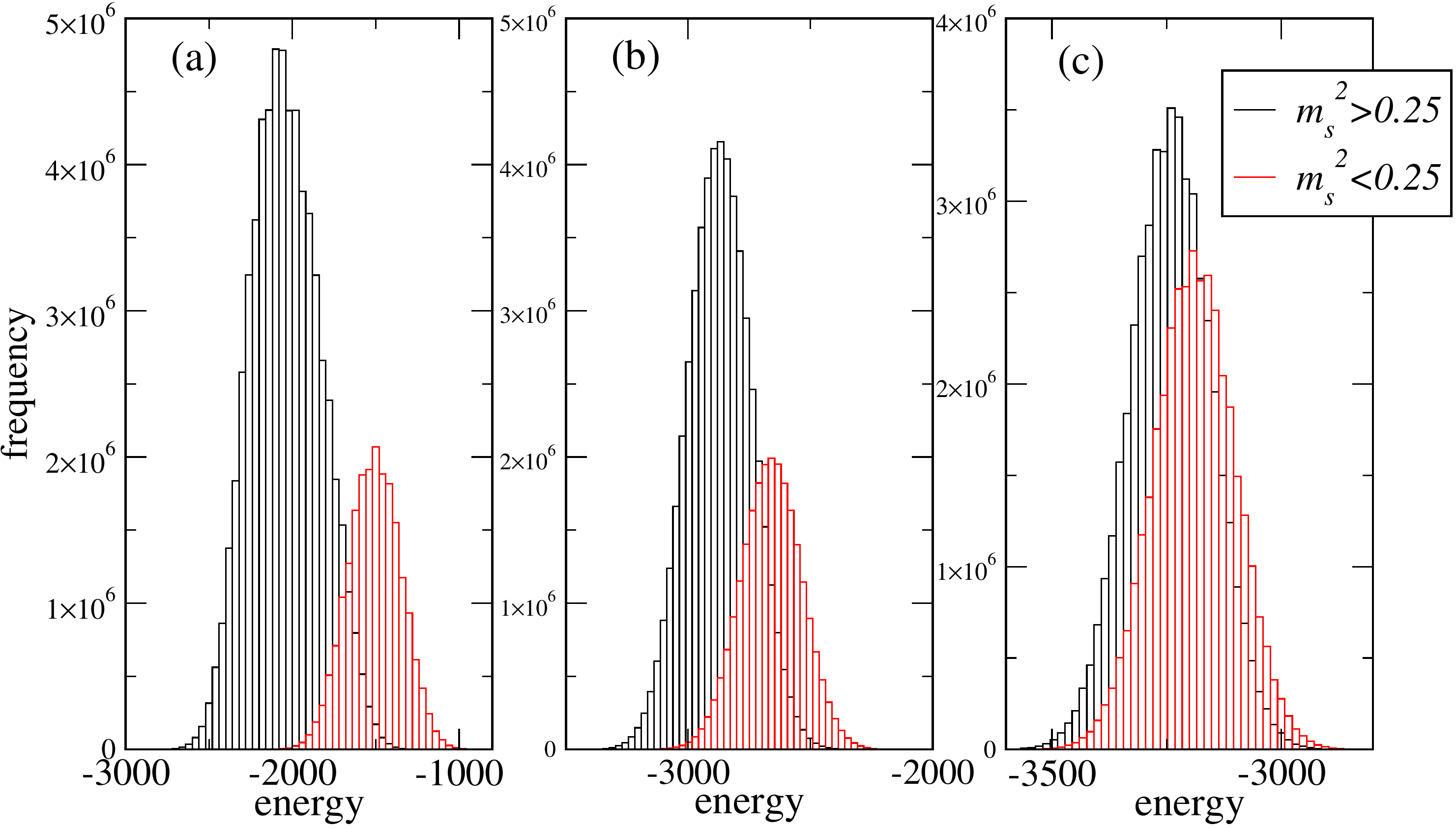}
\caption{(Color online) Energy histograms on $L=36$ cluster for (a) $h=0$, $T/J=2.274$; (b) $h/J=2$, $T/J=1.53$; (c) $h/J=2.35$, $T/J=0.86$. In each case, we have split data depending wheter $m_s^2$ is smaller or greater than 0.25.}
\label{fig:e0_histo}
\end{figure}

Concerning the energy distribution at the transition, we know from the analysis of its Binder cumulant $V_L$ that it is non gaussian. In fact, if we plot separately energy histograms obtained with configurations having $m_s^2$ smaller or greater than 0.25 (i.e. we separate contributions from both peaks in the $m_s^2$ histograms), then we observe in Fig.~\ref{fig:e0_histo} that data can be well represented by two gaussians centered at slightly different positions $E_\pm$. As explained in Ref.~\onlinecite{Challa1986}, at the transition the two gaussians have the same weight~\cite{note_histo}, leading to a dip in $V_L$ of size $2(E_+^4+E_-^4)/3(E_+^2+E_-^2)^2$. Therefore a first (respectively second) order transition will occur if $E_\pm$ are distinct in the thermodynamic limit (respectively if they merge). We have already performed this analysis in Fig.~\ref{fig:Cvmax_Vmin}(b) indicating that phase transitions for $h=0$ or $h=2J$ are second-order, while $h=2.35 J$ corresponds to first order. Looking at data on 
a single size $L=36$ does not give any indication since the peaks are \emph{more} separated for small $h$ where the transition is second-order, than for $h=2.35J$ (see Fig.~\ref{fig:e0_histo}). Clearly, this confirms that a careful finite-size study is necessary to ascertain the order of the phase transition.

\subsection{Phase diagram and discussion}
\label{conclud}

\begin{figure}
\includegraphics[width=\columnwidth,clip]{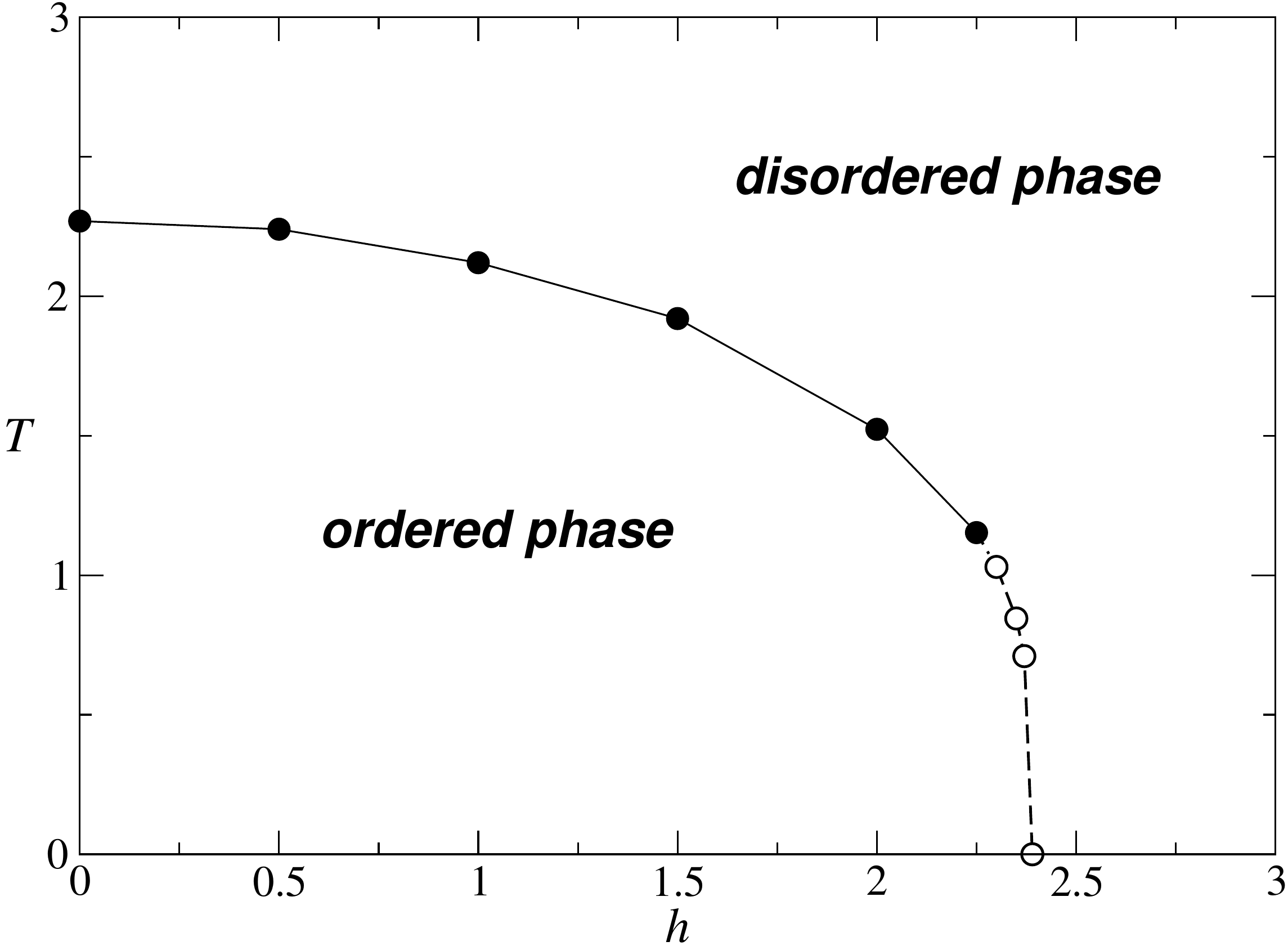}
\caption{(Color online) Phase diagram of the BWTF vs $(h,T)$. Full (respectively dashed) lines are used for second (respectively first) order phase transitions. Both lines meet at a putative tricritical point close to $h^*\simeq 2.3 J$ and $T^*\simeq J$.}
\label{fig:phasediag}
\end{figure}

In summary with this numerical study, we have been able to identify
clear signatures of second-order thermal phase transitions for various
transverse field amplitudes ranging from $h=0$ up to $h/J \sim 2.25$,
all occurring in the same universality class ($4$-state Potts model,
with no logarithmic corrections) as the classical BW model at
$h=0$. At the same time, the quantum critical point at $T=0$ (obtained
for $h_c (T=0)\simeq 2.4J$) is clearly of first-order nature. This
could be conjectured given that at $T=0$, the model would be in the
same universality class as the 4-state Potts model in 2+1 (or higher)
dimension, which is known to host a first-order transition. Given that
all correlation lengths are finite, we therefore expect that this
first-order character subsists at finite temperature, at least close
to $h_c(T=0)$. We have found indications that this is indeed the case
in our QMC simulations at $h=2.35J$. All results are summarized in the
phase diagram Fig.~\ref{fig:phasediag}.

We would thus expect a tricritical point at $(h^*,T^*)$ where the
first and second order transition lines meet (dots in Fig.~\ref{fig:phasediag}). Our best estimates are \mbox{$h^* \in ]2.25J,2.35J[$} and \mbox{$T^* \in ] 0.85J, 1.15J[$}. We are not aware of a theoretical prediction for the universality class of a tricritical point separating a Baxter-Wu universality class continuous transition line from a first-order line. With the current lattice sizes $L$ at hand in our QMC simulations, we have not been able to further characterize this multi-critical point (for instance computing precise critical exponents). While it is always difficult to extract critical exponents of tricritical points from finite-lattice simulations, the situation is particularly challenging for the BWTF model. There are indeed, on finite systems, signatures typical of first-order transitions (which vanish in the thermodynamic limit) even on the {\it continuous} transition line. Our current simulations close to the putative tricritical point (in the regime $]2.25J,2.35J[$) indicate that the correlation length (effective) critical exponent $\nu$ varies from $\nu=0.6$ at $h=2.25J$ to $\nu=0.57$ at $h=2.3J$ and then finally $\nu=0.52$ 
at $h=2.35J$ corresponding to first-order behavior.

\section{Conclusion}
\label{conclusion}

We have studied quantum and thermal phase transitions of the Baxter-Wu model in a transverse magnetic field using large-scale quantum Monte Carlo simulations and series expansions.
This has allowed us to characterize the full phase diagram vs $(h/J,T/J)$ (see Fig.~\ref{fig:phasediag}). On the one hand, our results confirm that the BWTF undergoes a first-order 
quantum phase transition at \mbox{($h\approx 2.4$, $T=0$)} that extends to finite temperature 
regime. On the other hand, the classical second-order phase transition at \mbox{($h=0$, $T\approx 2.226$)}, known to be in the 4-state Potts model universality class, also persists at finite magnetic field, up to rather large values of  $h \simeq 2.25 J$, with the same critical exponents. Therefore, we naturally predict the existence of a tricritical point located where these two phase boundaries merge, i.e.~approximately at ($h\approx 2.3J$, $T\approx J$), but its determination (including critical exponents) remains challenging. Indeed, even the second-order phase transition line exhibits typical signatures of first-order transitions such as negative Binder cumulant $U_L$ or double peaked energy histograms, which vanish in the thermodynamic limit but harden the analysis.

Given the central role played by tricritical points in Potts model physics~\cite{Nienhuis1979}, characterization of this new  tricriticality (critical exponents as well as determination of its conformal field theory) remains an exciting challenge that we hope to address in the future. One could also imagine studying the role of quantum fluctuations in other statistical models with multi-spin interactions such as the Hintermann-Merlini-Baxter-Wu generalizations on any plane Eulerian triangulation~\cite{Huang2013}. 
We also point to a recent reference~\cite{Velonakis2013} where the case of a \emph{classical} magnetic field was considered: a classical Monte-Carlo analysis indicate a different universality class ($\nu=1$, $\alpha=1/2$, $\beta=3/4$), however the critical exponents violate scaling relations so that further work should clarify this. Also, the latter situation breaks the sublattice symmetries that is present in our model so that there is no direct connection.

As stated above, the TCC in a parallel magnetic field 
on the honeycomb lattice is isospectral to the BWTF. The finite temperature error threshold 
of the TCC in the framework of the random 3-body Ising model \cite{katzgraber} as well as 
its zero-temperature robustness in the presence of a general uniform magnetic field 
or ferromagnetic Ising interactions \cite{ssj,ssj2} has already been studied. However, the
finite temperature physics of the TCC in a parallel magnetic field, is to the best of 
our knowledge not known. It is certainly also an interesting project to extend the 
investigation of the finite-temperature properties to the TCC in a parallel magnetic field.

\section{Acknowledgements}
We thank T. Roscilde and L. Turban for their useful comments on this work. 
SC and FA thank B. Eydoux for his collaboration in a related project. 
QMC simulations used the ALPS libraries~\cite{ALPS} and were performed using HPC resources from GENCI--CCRT, GENCI--IDRIS (Grant NO. x2013050225) and CALMIP.


\begin{thebibliography}
\bibitem{remark3} In order to accomodate all classical ground-states of the BW model, we only use $L\times L$ triangular cluster with $L$ multiple of 3.
\bibitem{note_histo} In this analysis, we have chosen $T_c(L)$ at the maximum slope of $m_s^2(T)$, which is not exactly the same as the temperature at which $V_L$ exhibits a dip. Therefore the energy histograms are not exactly of the same weight.
\end{thebibliography}
\end{document}